\documentclass[useAMS,usenatbib]{mn2e}
\usepackage{subfigure}
\usepackage{graphicx}

\title[A mid-infrared view of Mrk\,1066]{A mid-infrared view of the inner parsecs 
of the Seyfert galaxy Mrk\,1066 using CanariCam/GTC}
\author[C. Ramos Almeida et al.]
{\parbox{\textwidth}{C. Ramos Almeida$^{1,2}$\thanks{Marie Curie Fellow. E-mail: cra@iac.es},
A. Alonso-Herrero$^{3}$\thanks{Augusto Gonz\'alez Linares Senior Research Fellow},
P. Esquej$^{4}$,
O. Gonz\'alez-Mart\' in$^{1,2}$,
R. A. Riffel$^{5}$,
I. Garc\' ia-Bernete$^{1,2}$
J. M. Rodr\' iguez Espinosa$^{1,2}$,
C. Packham$^6$,
N. A. Levenson$^7$,
P. Roche$^8$,
T. D\' iaz-Santos$^9$,
I. Aretxaga$^{10}$,
C. \'Alvarez$^{1,2}$
}\vspace{0.4cm}\\
\parbox{\textwidth}{$^1$Instituto de Astrof\' isica de Canarias, Calle V\' ia L\'actea, s/n, E-38205, La Laguna, Tenerife, Spain\\
$^2$Departamento de Astrof\' isica, Universidad de La Laguna, E-38206, La Laguna, Tenerife, Spain\\
$^3$Instituto de F\' isica de Cantabria, CSIC-Universidad de Cantabria, E-39005, Santander, Spain\\
$^4$Departamento de Astrof\' isica, Facultad de CC. F\' isicas, Universidad Complutense de Madrid, E-28040, Madrid, Spain\\ 
$^5$Departamento de F\' isica, CCNE, Universidade Federal de Santa Maria, 97105-900 Santa Maria, RS, Brazil\\
$^6$Department of Physics and Astronomy, University of Texas at San Antonio, One UTSA Circle, San Antonio, USA\\
$^7$Gemini Observatory, Casilla 603, La Serena, Chile\\ 
$^8$Department of Physics, University of Oxford, Oxford OX1 3RH, UK\\
$^9$\emph{Spitzer} Science Center, California Institute of Technology, MS 220-6, Pasadena, CA 91125, USA\\
$^{10}$Instituto Nacional de Astrof\' isica, \'Optica y Electr\' onica (INAOE), 72000 Puebla, M\'exico
}}

\begin{document}

\date{}

\pagerange{\pageref{firstpage}--\pageref{lastpage}} \pubyear{2014}

\maketitle

\label{firstpage}

\begin{abstract}
We present mid-infrared (MIR) imaging and spectroscopic data of the Seyfert 2 galaxy Mrk\,1066 
obtained with CanariCam (CC) on the 10.4 m Gran Telescopio CANARIAS (GTC). The galaxy was observed in imaging mode 
with an angular resolution of 0.24\arcsec~(54 pc) in the Si-2 filter (8.7 \micron). The image reveals 
a series of star-forming knots within the central $\sim$400 pc, after subtracting the dominant active galactic nucleus (AGN) 
component. We also subtracted this AGN unresolved component from the 8--13 \micron~spectra of the knots and the nucleus,
and measured equivalent widths (EWs) of the 11.3 \micron~Polycyclic Aromatic Hydrocarbon (PAH) feature which are typical of pure starburst 
galaxies. This EW is larger in the nucleus than in the knots, 
confirming that, at least in the case of Mrk\,1066, the AGN dilutes, rather than destroys, the molecules responsible for the 
11.3 \micron~PAH emission. 
By comparing the nuclear GTC/CC spectrum with the \emph{Spitzer}/IRS spectrum of the galaxy, we find that the AGN component 
that dominates the continuum emission at $\lambda<$15 \micron~on scales of $\sim$60 pc (90--100\%) 
decreases to 35--50\% when the emission of the central $\sim$830 pc is considered. On the other hand, the AGN contribution
dominates the 15--25 \micron~emission (75\%) on the scales probed by \emph{Spitzer}/IRS.
We reproduced the nuclear infrared emission of the galaxy with clumpy torus models, and derived a torus gas mass of 
2$\times10^5~M_{\odot}$, contained in a clumpy structure of $\sim$2 pc radius and with a column density compatible with 
Mrk\,1066 being a Compton-thick candidate, in agreement with X-ray observations.
We find a good match between the MIR morphology of Mrk\,1066 and the extended Pa$\beta$, Br$\gamma$ and [O III]$\lambda$5007 
emission. This coincidence implies that the 8.7 \micron~emission is probing star formation, dust in the narrow-line region, 
and the oval structure previously detected in the near-infrared. On the other hand, the \emph{Chandra} soft X-ray morphology 
does not match any of the previous, contrary to what it is generally assumed for Seyfert galaxies. A thermal origin 
for the soft X-ray emission, rather than AGN photoionization, is suggested by the different data analyzed here.  
\end{abstract}


\section{Introduction}
\label{intro}

The interplay between nuclear activity and star formation in galaxies is still not well established. 
There is theoretical and observational evidence for active galactic nuclei (AGN) quenching star formation 
through the so-called AGN feedback (see e.g. \citealt{Granato04,Ho05,Springel05,Schawinski07,Schawinski09}), but 
the physical scales on which this quenching takes place, if it does, are not clear 
yet. Indeed, star formation is detected on kpc-scales down to tens of parsecs from the AGN in Seyfert galaxies 
\citep{Diamond10,Diamond12,LaMassa12,Alonso14,Esquej14}, and numerical simulations predict a relation with 
some scatter between the star formation rates (SFRs) 
on different galaxy scales (between 10 kpc and 1 pc) and the black hole accretion rates \citep{Hopkins10}.


Ground-based mid-infrared (MIR) observations with 8--10 m-class telescopes might hold the key 
for disentangling the relation between nuclear activity and star formation on parsec-scales.  
The limited spatial resolution of the \emph{Spitzer} Space Telescope (hereafter {\it \emph{Spitzer}}; 
$\sim$4--5\arcsec) only allows to study this relation on kpc-scales, 
although with extremely good sensitivity. Therefore, we are conducting a MIR imaging and spectroscopic 
survey of $\sim$100 local AGN using the instrument CanariCam (CC; \citealt{Telesco03}) on the 10.4 m Gran Telescopio 
CANARIAS (GTC), in La Palma. The sample includes both high-to-intermediate luminosity AGN (PG quasars, radio galaxies
and Seyfert galaxies) and low-luminosity AGN (low-ionization nuclear emission-line regions; LINERs) covering
almost six orders of magnitude in AGN luminosity (see \citealt{Alonso13} for further details).    


Among the galaxies in our survey already observed with GTC/CC, we selected the Seyfert 2 (Sy2) 
Mrk\,1066 (UGC\,2456), for which there is evidence in the literature of circumnuclear star formation 
as well as extended NIR line emission (e.g. Pa$\beta$, Br$\gamma$, [Fe II]; \citealt{Riffel10}). This barred galaxy, at a distance of 
47.2 Mpc, has been extensively studied at different wavelengths \citep{Bower95,Ramos09,Smirnova10,Riffel10,Riffel11}. Our 
goal is to understand the different mechanisms responsible for the MIR emission in the central kpc of Mrk\,1066, including emission 
from the dusty torus, dust emission from the narrow-line region (NLR) and star formation.  
In \citet{Alonso14} we present a complementary, detailed study of the extended PAH emission of five local AGN, 
including Mrk\,1066, using GTC/CC spectroscopy. The latter work focuses on star formation only, and seeks for general 
results on PAH destruction/dilution by the intense AGN continuum.


The presence of on-going star formation in the central kiloparsec of 
Mrk\,1066 was clearly revealed by integral field spectroscopic data from the Near-Infrared Integral 
Field Spectrometer (NIFS) on the 8 m Gemini North telescope. The data was obtained 
using adaptive optics (AO), which provided an angular resolution of $\sim$0.1--0.2\arcsec~($\sim$35 pc; \citealt{Riffel10,Riffel11}), and
showed line-emitting gas elongated in the same direction of the [O III] and the radio 
emission (PA=315$\degr$). From the NIR emission line kinematics, 
\citet{Riffel11} reported the existence of a compact rotating disc of $\sim$70 pc radius and an outflow with the same 
orientation as the radio jet.

The emission-line ratios measured by \citet{Riffel10} along the ionization cones are typical of 
Seyfert galaxies, indicating that the active nucleus is the dominant source of ionization in the NLR, with 
some contribution from shock excitation. 
On the other hand, away from the ionization cones the line ratios are characteristic of 
star-forming regions. These findings explain the results from optical and near-infrared (NIR) data at lower
spatial resolution \citep{Bower95,Ramos09}, in which we observe a mixture of high and low ionization regions.



Mrk\,1066 then offers the perfect scenario for studying the interplay between nuclear activity and star formation 
in the inner $\sim$400 pc of the galaxy, by comparing our new GTC/CC imaging and spectroscopic observations with available
multifrequency data of similar resolution. 

Throughout this paper we assume a cosmology with H$_0$ = 73 km s$^{-1}$ Mpc$^{-1}$, $\Omega_m$ = 0.27, 
and $\Omega_{\Lambda}$ =0.73.

\section{Observations and data reduction}
\label{observations}

\subsection{GTC/CC imaging and spectroscopy}

The Sy2 galaxy Mrk\,1066 was observed in August 2013 with the MIR camera/spectrograph CanariCam (CC).
CC uses a Raytheon 320$\times$240 
Si:As detector which covers a field--of--view (FOV) of 26\arcsec$\times$19\arcsec~on the sky and its 
pixel scale is 0.0798\arcsec~(hereafter 0.08\arcsec). The standard MIR chopping--nodding technique was used to remove the time-variable sky
background, the thermal emission from the telescope and the detector 1/f noise, where f is the frequency of the noise component. 
The employed chopping and nodding throws were 15\arcsec, with chop and nod position angles of 45 and -135 deg respectively. 

Both imaging and spectroscopic observations 
were done in queue mode, on different nights 
and under photometric conditions. The data were taken as part of an ESO/GTC large programme (182.B-2005) awarded 180 hours 
of GTC/CC time, aimed to conduct a MIR survey of $\sim$100 
nearby AGN by exploiting the unique capabilities of CC on the GTC. See \citet{Alonso13} for a more detailed description 
of this MIR survey. 

MIR imaging observations of Mrk\,1066 were taken on 2013 Aug 27 with GTC/CC, using 
the narrow Si-2 filter ($\lambda_c$=8.7 \micron, $\Delta\lambda$=1.1 \micron). We took three exposures 
of 139 s each, which we combined once reduced to produce a single image of 417 s on-source integration time. The 
airmass during the observations was $\sim$1.2. Images in the same filter of the Point Spread Function (PSF) standard 
star HD 18449 were obtained immediately after the science target for accurately sampling the image quality, 
and to allow flux calibration. We measured an angular resolution of 0.24\arcsec~(54 pc) from the FWHM of the observed 
PSF standard star. 

\begin{figure}
\includegraphics[width=9cm]{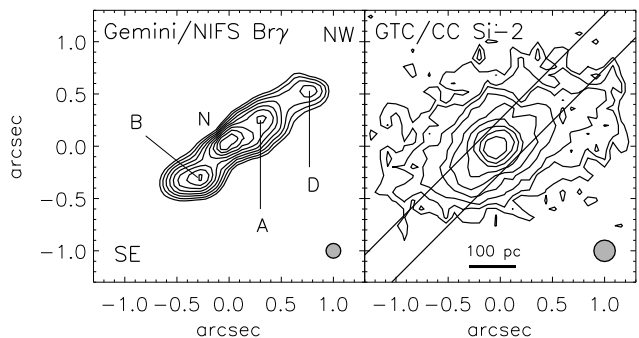}
\caption{
Left: Gemini/NIFS 2.166 \micron~image contours at the 3$\sigma$ level of Mrk\,1066 from \citet{Riffel10}, 
with the position of the active nucleus (N) and the three IR regions detected (A, B and D) indicated. 
Right: GTC/CC 8.7 \micron~image contours at the 3$\sigma$ level, with the position of the 0.52\arcsec~wide slit superimposed.
Both images have been smoothed using a Gaussian function with $\sigma$=0.5 pixels. 
Filled circles indicate the approximate angular resolution of the images 
(0.18\arcsec~and 0.24\arcsec, respectively).
\label{fig1}}
\end{figure}

For the spectroscopy, we employed the low spectral resolution GTC/CC N-band grating, with nominal resolution 
R = $\lambda$/$\Delta\lambda\sim$175 and covering the spectral range 7.5--13 \micron. The data were taken on 
2013 Aug 31 using the 0.52\arcsec~wide slit, oriented 
at 315$\degr$ to make it coincide with the axis of the ionization cones and the radio jet \citep{Bower95,Nagar99}, 
and passing through the nucleus and the A, B and D knots detected in the NIR (\citealt{Riffel10};
see left panel of Figure \ref{fig1}).


We first took an acquisition image in the Si-2 filter to ensure optimal placement of the slit, and after 
introducing the grating, we integrated for 1061 s on-source at an airmass of $\sim$1.1. Immediately after 
the observation of Mrk\,1066, we obtained a spectrum of the standard 
star HD 18449 to provide flux calibration and telluric and slit--loss corrections. Using the acquisition images of 
both the standard star and galaxy nucleus, we measured a FWHM of 0.26\arcsec~(58 pc). Thus, the nuclear size 
is consistent with an unresolved source, although with fainter extended emission.

We reduced the data using the {\it RedCan} pipeline for the reduction and analysis of MIR imaging and 
spectroscopic data \citep{Gonzalez13}. In the first step, {\it RedCan} uses keywords in the fit headers 
to identify the type of each observing block, and they are used accordingly throughout the pipeline. 

In the case of the imaging,
{\it RedCan} uses the Gemini IRAF packages\footnote{The released version of the Gemini IRAF package is an external 
package layered upon IRAF and is available to users and other interested parties 
(http://www.gemini.edu/sciops/data--and--results/processing--software).}, which include sky subtraction, 
stacking of the individual images and rejection of bad frames. {\it RedCan} also provides flux calibrated images 
when associated standard stars are observed, as it is the case for Mrk\,1066. The contours of the 
fully-reduced GTC/CC 8.7 \micron~image of Mrk\,1066 are shown in the right panel of Figure \ref{fig1}. 

For reducing MIR spectra, {\it RedCan} follows standard MIR reduction recipes, 
including sky subtraction, stacking of individual observations, rejection of bad frames, wavelength calibration, 
trace determination and spectral extraction. The latter step can be done either as point source or extended source. In 
the first case, {\it RedCan} uses an extraction aperture that increases with wavelength to take care of the
decreasing angular resolution, and it also performs a correction to account for slit losses. If the spectrum is 
extracted as an extended source, a fixed aperture is used and no slit--loss corrections are applied. In this work, 
we use both the GTC/CC nuclear spectrum extracted as point source and the spectra of the knots (labelled as 
N, A, B and D in Figure \ref{fig1}), extracted as extended sources (see Sections \ref{nuclear_spectroscopy} and 
\ref{extended_spectroscopy}). {\it RedCan} finally produces flux-calibrated spectra, which are combined in a single
one for each target.


\subsection{\emph{Chandra} imaging data}
\label{chandra}

We compiled \emph{Chandra} data of Mrk\,1066, taken with the Chandra Advanced CCD Imaging Spectrometer (ACIS) 
on 2003 July 14 (ObsID 4075). The data were reduced from level 2 event files using the CXC \emph{Chandra}
Interactive Analysis of Observations (CIAO) software version 4.4. Periods of
high background were removed from the observation using the task {\sc lc$_-$clean.cl}
in a source-free region of the sky of the same observation. The net exposure time
and net total number of counts in the 0.2-10 keV band are 20 ksec and 800 counts
respectively.

\emph{Chandra} data include information on the position where the photons fall into
the detector better than the one used with the default pixel size (i.e. 0.492\arcsec). Thus,
smaller spatial scales are accesible as the image moves across the detector pixel during
the telescope dither. This allowed us to sub-pixel binning our images to a pixel size of
0.06\arcsec. We extracted two images in the 0.5--2 and 2--10 keV bands and used the adaptive
smoothing techniques {\sc asmooth} to enhance weak structures \citep{Ebeling2006}. This
technique is particularly useful for images containing multi-scale complex structures, preserving
its spatial signatures. To do that, we selected a minimum and maximum significance S/N level of 1.5 and 3
respectively and a maximum scale of 2 pixels.

\subsection{Hubble Space Telescope imaging data}
\label{hst}

We downloaded a reduced [O III]$\lambda$5007 \AA~image from the \emph{Hubble} Space Telescope (\emph{HST}) science archive, 
taken with 
the Wide Field Planetary Camera (WFPC) on 1992 November 2, as part of \emph{HST} proposal 3724. The data were first published
by \citet{Bower95}. The galaxy was observed in the F492M and F547M filters, chosen to isolate the [O III]+H$\beta$ emission 
and their adjacent continuum, respectively. Using long-slit spectroscopy, \citet{Bower95} estimated that the [O III] emission
accounts for $\sim$84\% of the total emission-line flux in the F492M filter. The pixel scale of the WFPC detector is 
0.043\arcsec. 

\section{Results}
\subsection{The IR morphology of Mrk\,1066}
\label{imaging}

We employed the PSF star, which is a Cohen standard, observed after the 8.7 \micron~image of Mrk\,1066 in the same filter,
to determine the unresolved (i.e. nuclear) component of Mrk\,1066. This is done by scaling the maximum of the PSF star emission to 
the peak of the galaxy emission, and then integrating all the flux. By doing this,  
we obtained a nuclear (unresolved) 8.7 \micron~flux of 63$\pm$9 mJy. We estimated a total uncertainty of 15\%
by quadratically adding the errors in the flux calibration and point source extraction. 

Apart from allowing us to determine accurate nuclear fluxes, we use PSF subtraction to study the galaxy's extended emission,
after removing the dominant AGN component. We require a flat profile in the residual of the total emission minus the scaled
PSF for a realistic, and not over-subtracted, galaxy profile. The residual profiles from the different scalings we tried 
evidence the best-fitting result, which in this case is 100\%. See \citet{Radomski02} and \citet{Ramos09b} for further details. 

\begin{figure}
\includegraphics[width=9cm]{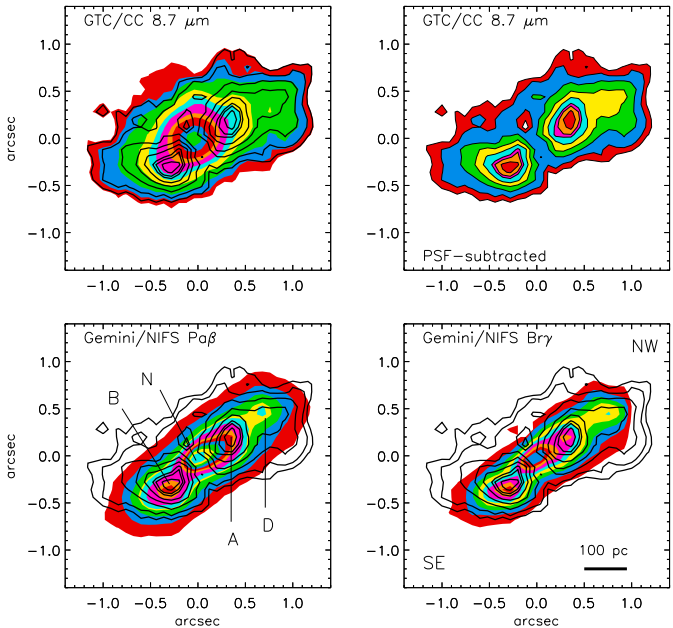}
\caption{Top: GTC/CC 8.7 \micron~and PSF-subtracted 8.7 \micron~image contours at the 3$\sigma$ level of Mrk\,1066.
Bottom: Gemini/NIFS Pa$\beta$ and Br$\gamma$ contours at the 3$\sigma$ level of Mrk\,1066 from \citet{Riffel10}, 
but interpolated to the pixel size of the GTC/CC images (0.08\arcsec). Black contours are those of the PSF-subtracted 8.7 \micron~image. 
North is up, and east to the left. 
\label{fig2}}
\end{figure}

In Figure \ref{fig2} we show the GTC/CC 8.7 \micron~image contours of Mrk\,1066 before (top left panel) and after PSF subtraction 
(top right panel). The PSF-subtracted image reveals a series of MIR knots (labelled in Figure \ref{fig1} as B, A and D, 
from SE to NW), which coincide with the position of those seen in the NIR using AO \citep{Riffel10}. 
In Table \ref{tab1} we report the positions of the MIR knots, as measured from the active nucleus (N), and their
8.7 \micron~fluxes, calculated as described above for the nucleus (PSF-subtraction) and in apertures of 0.25\arcsec~diameter for 
knots A, B and D (from the PSF-subtracted image). 

\begin{table}
\centering
\begin{tabular}{ccccc}
\hline
\hline
Knot & (X,Y) & \multicolumn{2}{c}{Distance} & 8.7 \micron~flux  \\
 & (arcsec) & (arcsec) & (pc) & (mJy) \\
\hline
N  &  (0.00,0.00)   &  \dots & \dots & 63$\pm$9 \\  
A  & (-0.38,-0.20)  &  0.43 & 100 & 3.3$\pm$0.5 \\ 
B  &  (0.27,0.30)   &  0.40 &  93 & 3.2$\pm$0.5 \\
D  & (-0.79,-0.40)  &  0.89 & 207 & 2.3$\pm$0.3 \\  
\hline
\end{tabular}
\caption{Positions, projected distances (measured from the nucleus) and  
fluxes of the four IR regions labelled in 
Figure \ref{fig1}. Fluxes of knots A, B and D 
were obtained in 0.25\arcsec~diameter apertures. 
The nuclear (unresolved) flux of knot N was
obtained from PSF-subtraction, as described in Section \ref{imaging}.}
\label{tab1}
\end{table}

In the bottom panels of Figure \ref{fig2} we show the flux map contours of Pa$\beta$ and Br$\gamma$ from 
\citet{Riffel10}. These maps were obtained with the integral field unit of Gemini-North NIFS and using AO. 
NIFS has a FOV of 3\arcsec$\times$3\arcsec~and a pixel size of 0.05\arcsec~was used to construct the flux maps 
in \citet{Riffel10}, which here we interpolated to match the pixel size (0.08\arcsec) of GTC/CC. 
The total exposure time was 4800 s in both the J and K bands (Pa$\beta$ and Br$\gamma$ respectively).

The Gemini/NIFS integral field data was employed by \citet{Riffel10} to measure emission-line ratios in each 
resolution element, as shown in their figure 8. Using the NIR ratios [Fe II]/Pa$\beta$ and H$_2$/Br$\gamma$, 
\citet{Riffel10} claimed that the gas in knots B and D
is predominantly photoionized by starbursts, whereas the nucleus and knot A have typical values of Seyfert galaxies. 
The NIR spectrum of knot A shown in figure 1 of \citet{Riffel10} is indeed very similar to the nuclear spectrum, 
both extracted in apertures of 0.25\arcsec$\times$0.25\arcsec. 

\begin{figure*}
{\par\includegraphics[width=5.8cm]{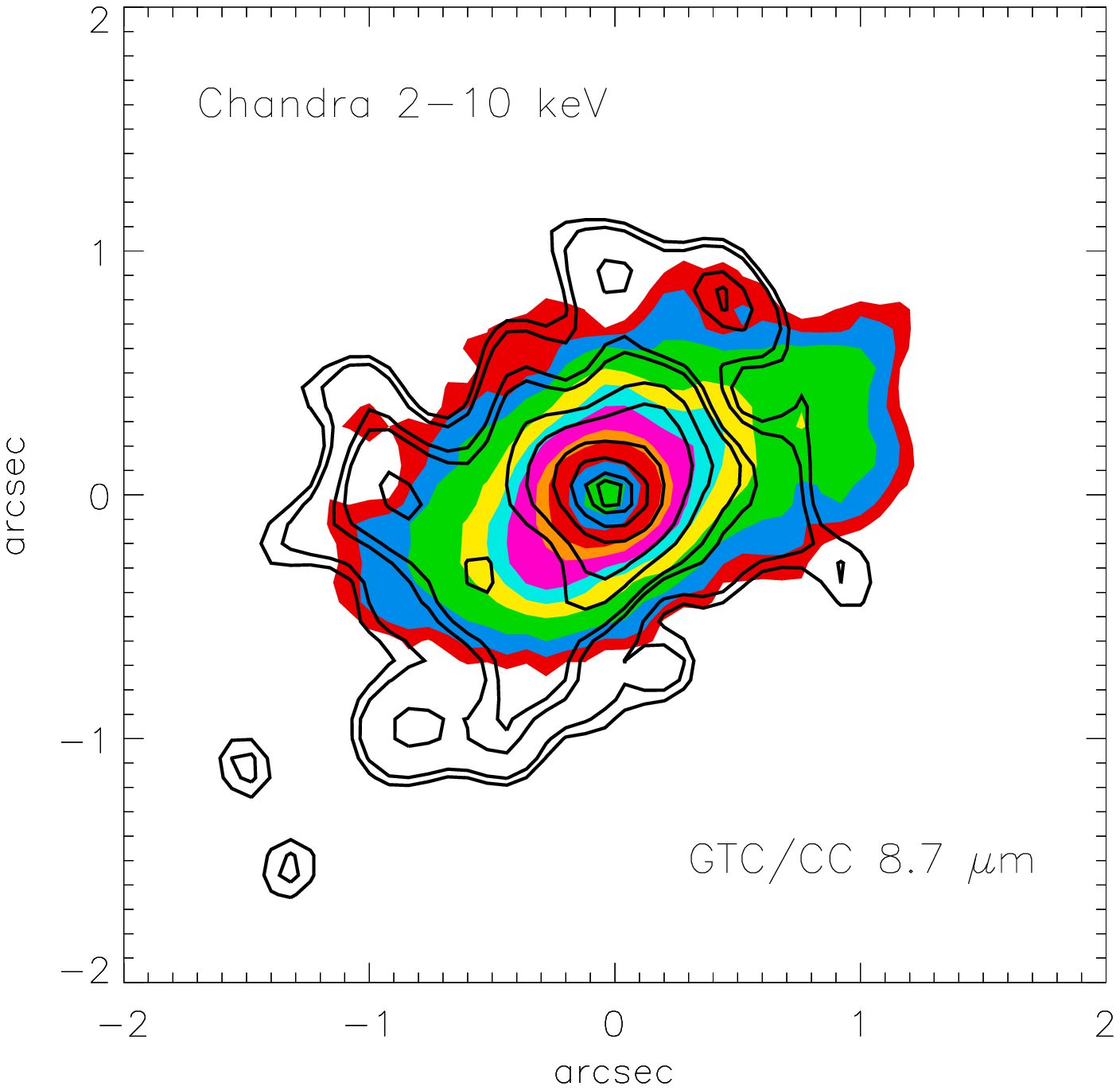}
\includegraphics[width=5.8cm]{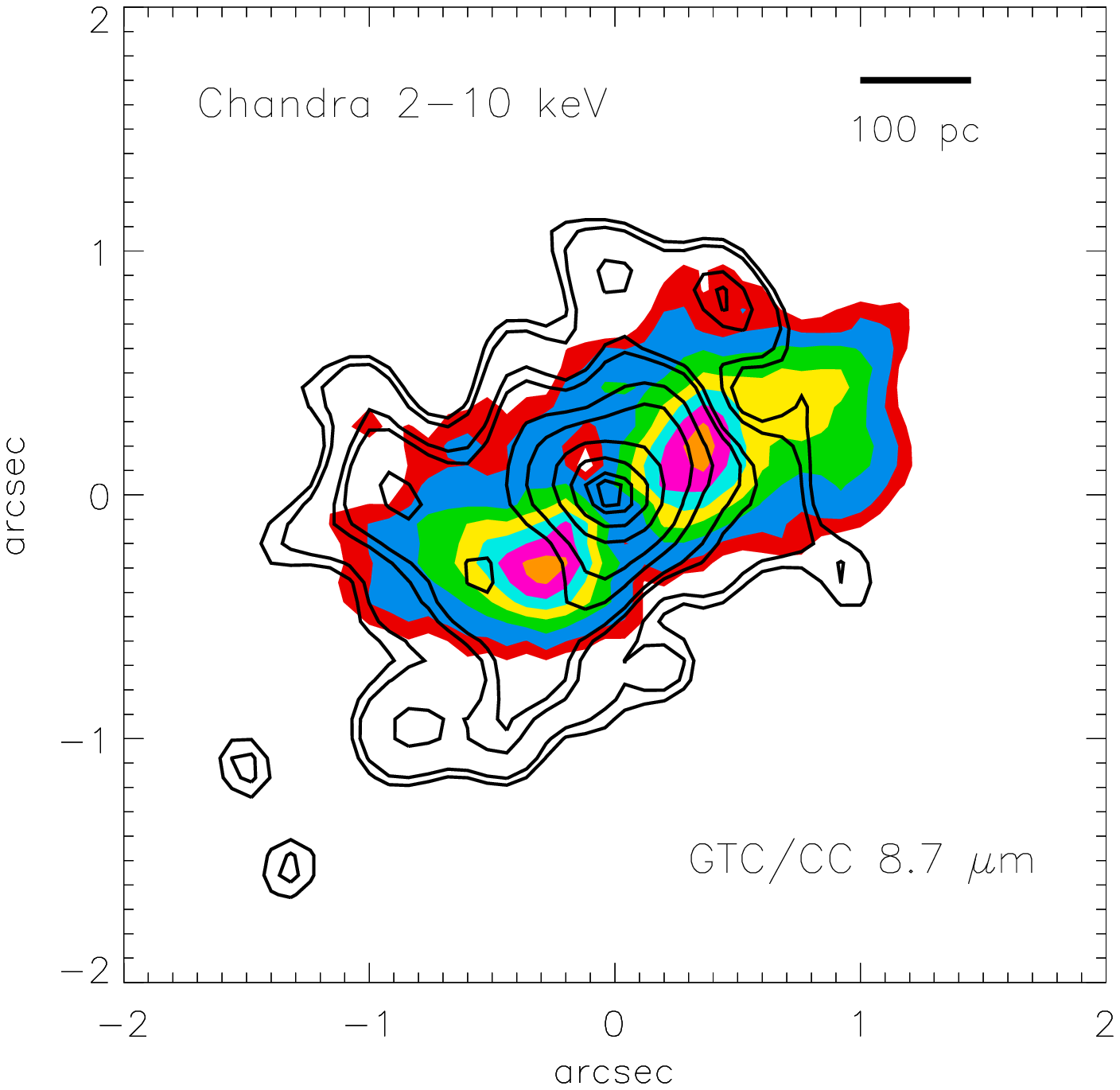}
\includegraphics[width=5.8cm]{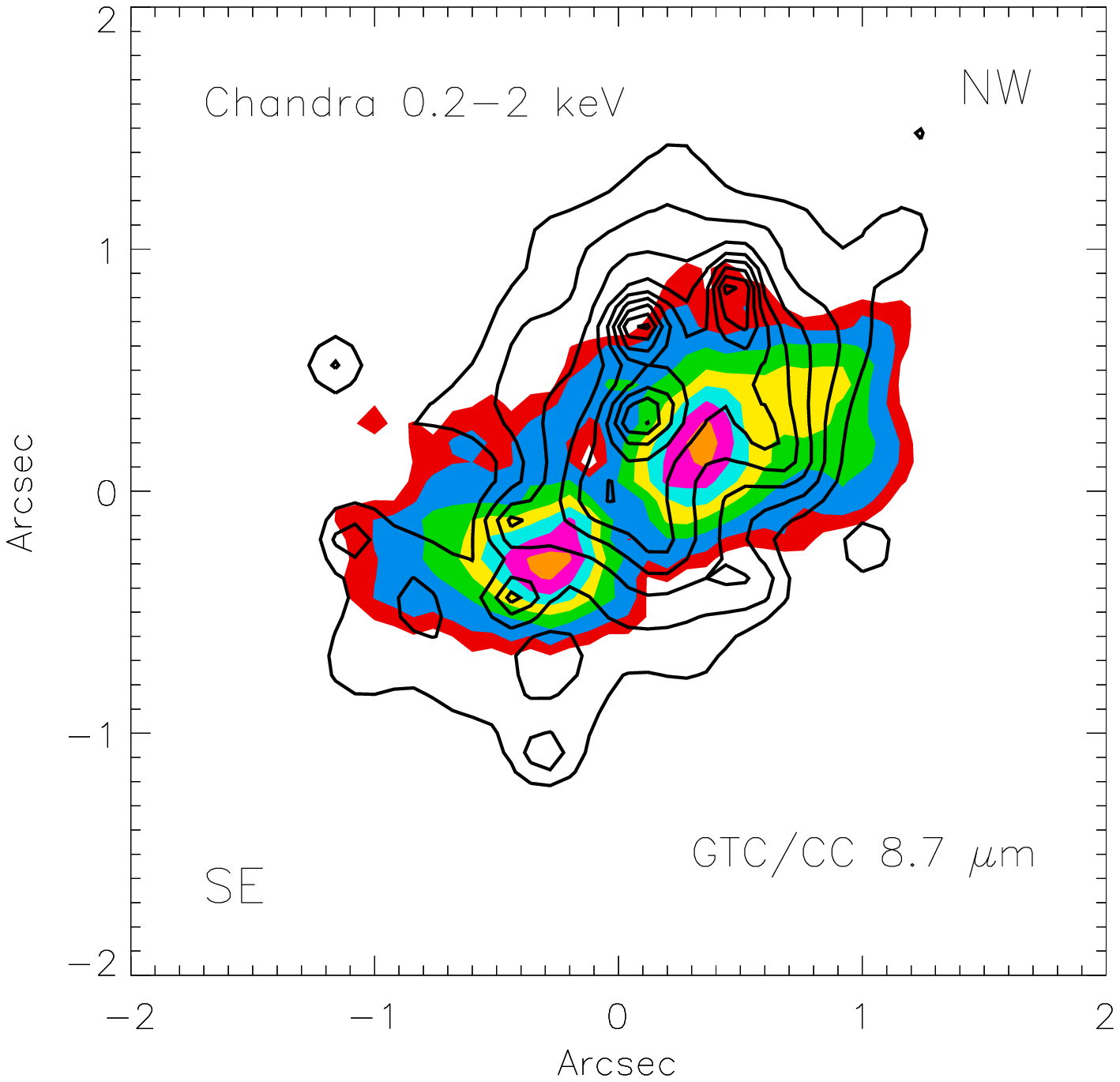}\par}
\caption{Left and center: contours, at the 3$\sigma$ level, of the hard (2--10 keV) X-ray emission of Mrk\,1066 (black lines), 
overlaid on the GTC/CC 8.7 \micron~contours (in colour) before and after subtracting the PSF. 
Right: same as in the central panel, but with the contours of the soft (0.5--2 keV) X-ray emission overlaid.
The \emph{Chandra}/ACIS images were interpolated to the pixel size of GTC/CC (0.08\arcsec). 
\label{fig8}}
\end{figure*}

In Figure \ref{fig8} we show the \emph{Chandra} contours of the hard (2--10 keV) and soft (0.5--2 keV) X-ray emission, once
interpolated to the pixel size of the GTC/CC image (0.08\arcsec), overlaid on the GTC/CC Si-2 image of 
Mrk\,1066. The left panel of Figure \ref{fig8} reveals a single nucleus emitting in hard X-rays and coinciding with the 
position of the MIR nucleus. Besides, the hard X-ray contours 
do not match the MIR morphology when we subtract the PSF component, as shown in the central panel of Figure \ref{fig8}. 
Thus, we find that the Seyfert-like emission reported by \citet{Riffel10} for knot A is not associated with 
a hard X-ray emitting source, as it is the MIR nucleus. 

The soft X-ray emission of Mrk\,1066, on the other hand, is elongated along a different axis, 
with a slightly larger PA than the NIR and MIR 
emission. In addition, it shows three knots towards the NW, which do not coincide with any of the IR knots (see right
panel of Figure \ref{fig8}). 
The soft X-ray knots have 2--10 keV luminosities of 1.1, 1.2 and 1.9$\times 10^{39}~erg~s^{-1}$, as measured from the 
\emph{Chandra} data, the NW knot
being the most luminous. These luminosities are in the boundary between those of 
luminous X-ray binaries and ultra-luminous X-ray sources (ULXs; L$_X\sim1\times 10^{39}~erg~s^{-1}$, \citealt{Miller04,Fabbiano06,
Gonzalez06}).

Considering the availability of multiwavelength
archival data publicly available for Mrk\,1066, of similar angular resolution, we use them to assess the 
importance of recent star formation activity, NLR dust and shocks in the inner $\sim$400 pc of the galaxy. This 
is discussed in detail in Section \ref{discussion1}.

\subsection{Nuclear and extended IR emission}
\label{nuclear_spectroscopy}

\begin{figure*}
\par{
\includegraphics[width=8.6cm]{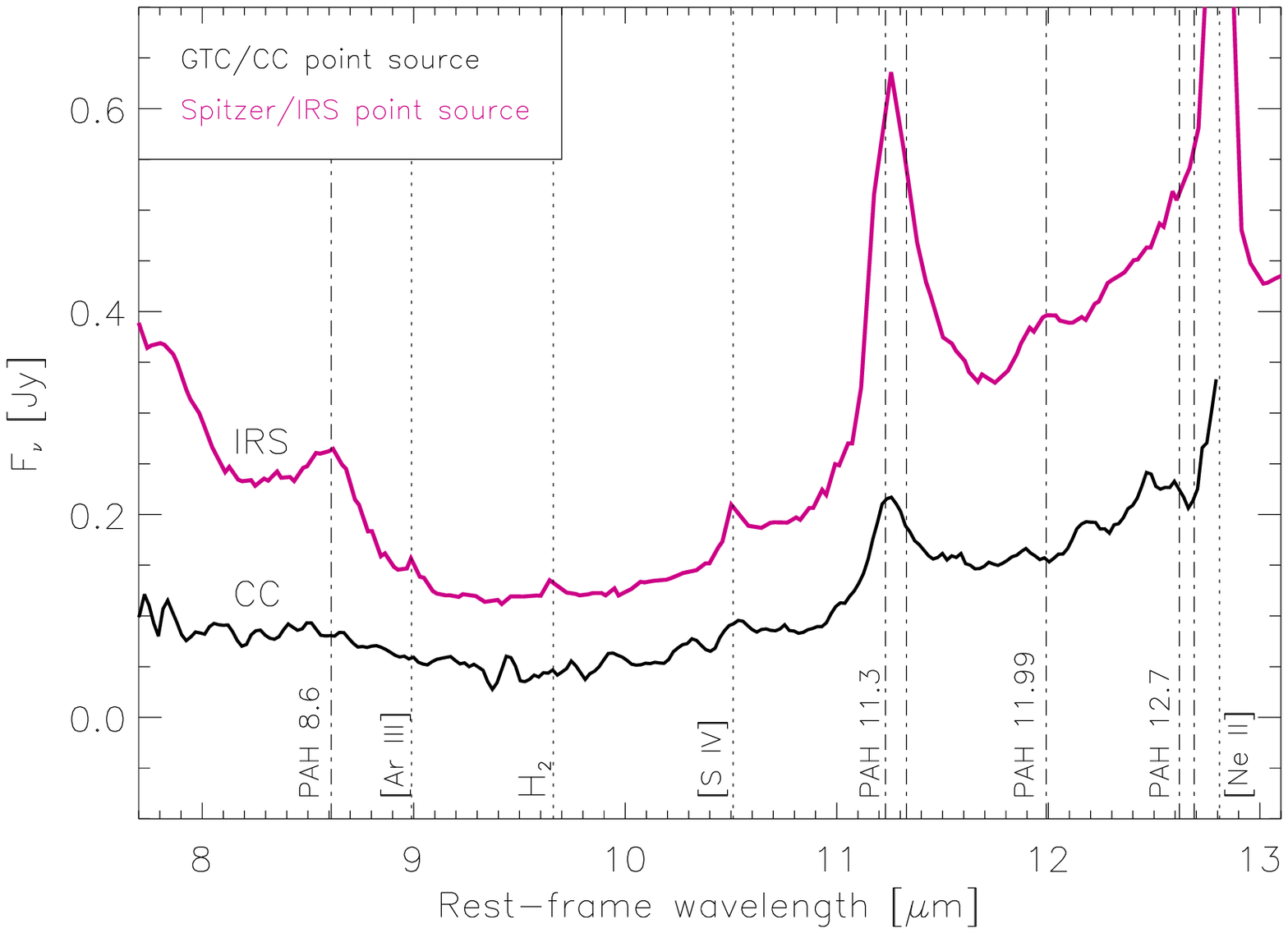}
\includegraphics[width=8.6cm]{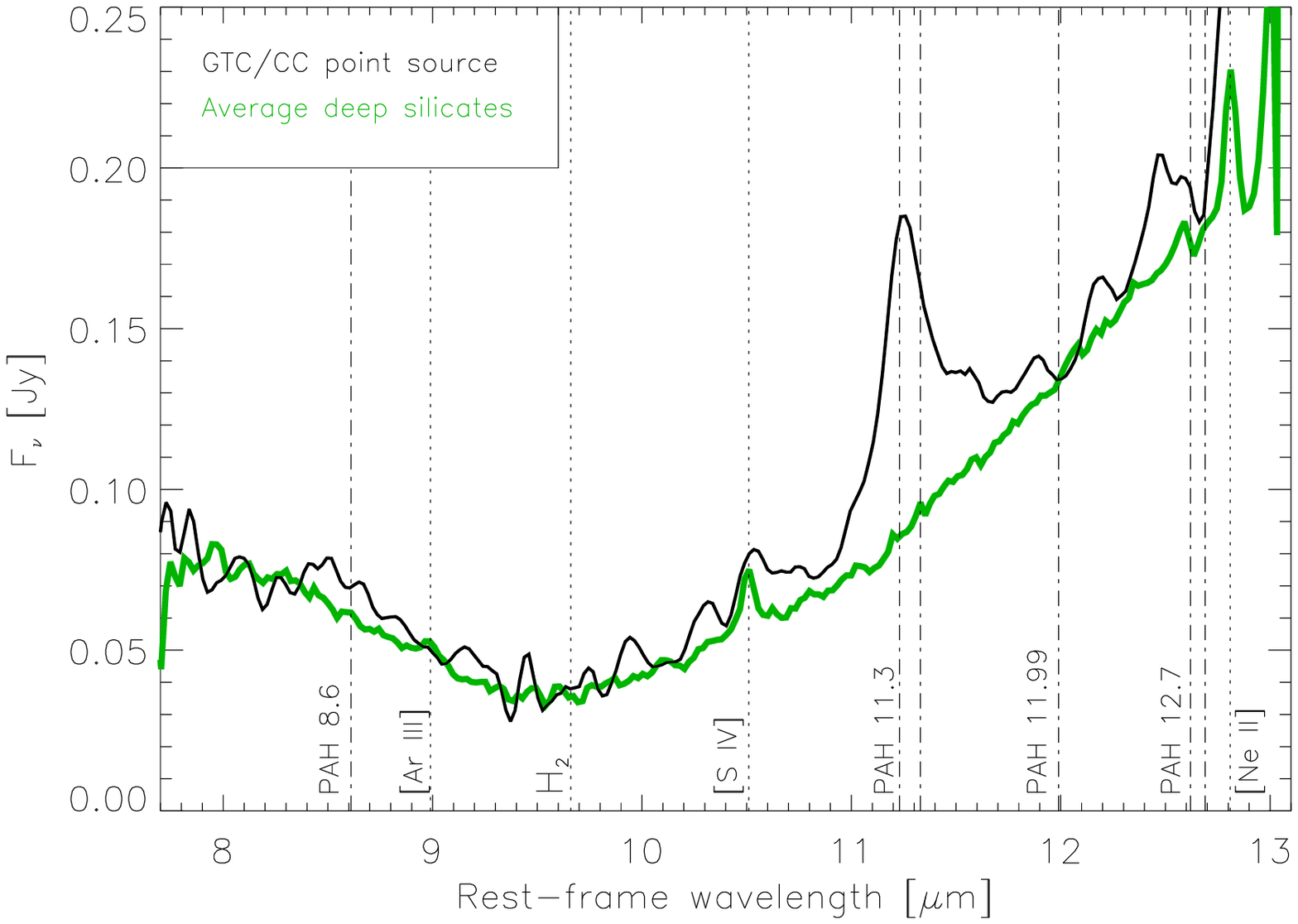}\par}
\caption{Left: Nuclear GTC/CC N-band and IRS/\emph{Spitzer} spectra (black and pink lines respectively) of 
Mrk\,1066. Both spectra have been extracted as point sources, which corresponds to the central $\sim$60 and $\sim$830 pc
respectively. Right: Comparison between the nuclear GTC/CC 
spectrum and the stacked nuclear spectrum of the seven galaxies with relatively deep silicate features in \citet{Esquej14}, scaled
at 12 \micron.
In both panels, the vertical dotted lines indicate the position of typical star-forming regions/AGN emission lines/bands. 
\label{fig3}}
\end{figure*}

In Figure \ref{fig3} we show the nuclear spectrum of Mrk\,1066, extracted as 
a point source. There is a good 
agreement between the flux calibration of the nuclear spectrum provided by {\it RedCan} and the nuclear 8.7 \micron~flux obtained from the 
imaging in Section \ref{imaging}, as we only measured a 15\% mismatch between the two.  
For consistency, we scaled the spectrum to the nuclear flux obtained from the CC image and 
we estimated a 15\% total uncertainty for the CC spectrum by quadratically adding the errors in the flux calibration 
and point source extraction.

The most intense feature in the GTC/CC nuclear spectrum shown in the left panel of Figure \ref{fig3}
is the 11.3 \micron~PAH band, which reveals the presence of 
star formation in the central $\sim$60 pc. The 8.6 \micron~PAH feature is weaker, although it becomes more
conspicuous in the 0.52\arcsec$\times$2\arcsec~spectrum shown in figure 1 in \citet{Alonso14}. 

For comparison, in the left panel of Figure \ref{fig3} we show the \emph{Spitzer}/Infrared Spectrograph (IRS) spectrum, retrieved from 
Cornell Atlas of \emph{Spitzer}/IRS Source (CASSIS v4; \citealt{Lebouteiller11}). The spectrum was obtained in staring mode
with the short-low (SL) module, which covers the range $\sim$5-15 \micron~and provides spectral resolution R = 60--120. 
Considering the typical angular resolution of 3.7\arcsec~for the SL module, given by the slit width, this corresponds 
to a physical scale of 830 pc. This is 14 times larger than the physical scale probed by the CC spectrum, of $\sim$60 pc.
The \emph{Spitzer}/IRS spectrum shows prominent 8.6 and 11.3 \micron~PAH features, as well as intense [Ne II]$\lambda$12.81 \micron~and 
relatively weak [S IV]$\lambda$10.51 \micron~(see left panel of Figure \ref{fig3}). 

In the right panel of Figure \ref{fig3} we compare the nuclear GTC/CC spectrum of Mrk\,1066 and the stacked spectrum of the 
seven galaxies with relatively deep silicate features in \citet{Esquej14}. This stacking was done using nuclear 
N-band spectra from the Gemini instruments T-ReCS and MICHELLE of Sy2 galaxies in the RSA sample, which
also probe spatial scales of $\sim$60--65 pc. From this comparison, it is clear that 
the 9.7 \micron~silicate feature that we detect in the GTC/CC nuclear spectrum of Mrk\,1066 is among the deepest  
reported in \citet{Esquej14}. By performing a linear fitting of its adjacent continuum (using the two featureless regions
8--8.2 \micron~and 12--12.2 \micron), we measured an apparent optical depth of the silicate feature
$\tau_{9.7}$=1.11. This value is among the largest reported for a sample of nearby Seyfert galaxies by \citet{Gonzalez13},
measured from Gemini/T-ReCS data. In particular, it is coincident with the $\tau_{9.7}$ values measured for NGC\,3281, NGC\,5506 and NGC\,7582, 
which are edge-on galaxies with nuclear dust lanes. Mrk\,1066 has an intermediate inclination, but its foreground extinction in the nucleus 
is also high (A$_V$$\sim$5 mag; \citealt{Riffel10}) and could be contributing to the silicate absorption.

\subsubsection{Nuclear SED modelling with clumpy torus models}
\label{clumpy}

Recent success in explaining several properties of the nuclear IR spectral energy distributions (SEDs) of 
Seyfert galaxies has been gathered under the assumption of a clumpy distribution of dust surrounding AGN
\citep{Mason09,Nikutta09,Ramos09b,Ramos11b,Ramos11c,Honig10b,Alonso11,Alonso12,Alonso13,Lira13}.


{\it BayesClumpy}\footnote{https://github.com/aasensio/bayesclumpy} \citep{Asensio09,Asensio13} is a computer program 
that can be used to fit photometry and/or spectra with the clumpy dusty torus models of \citet{Nenkova08a,Nenkova08b}. 
The fitting is done in a Bayesian scheme, carrying out inference over the model parameters 
for observed SEDs. Therefore we can specify a-priori information about the model
parameters. We consider the priors to be truncated uniform distributions for the six model parameters 
in the intervals reported in Table \ref{tab2}. 
See \citet{Ramos14} for a detailed description of the model parameters 
and examples of SED fitting with {\it BayesClumpy}.


\begin{table}
\centering
\begin{tabular}{cllc}
\hline
\hline
Parameter &  Prior & Median & MAP \\
\hline
$\sigma$	 	 & [15\degr, 75\degr]	   & 62\degr$\pm^{5}_{7}$	& 68\degr    	\\
$Y$			 & [5, 100]		   & 33$\pm^{31}_{10}$  	& 19		\\
$N_0$			 & [1, 15]		   & 12$\pm$2    		& 13		\\
$q$		 	 & [0, 3]		   & 0.8$\pm^{0.6}_{0.5}$	& 0.2		\\
$i$		 	 & [0\degr, 90\degr]	   & 35\degr$\pm^{21}_{12}$	& 17\degr    	\\
$\tau_{V}$		 & [5, 150]		   & 66$\pm^{10}_{15}$  	& 72		\\ 
$A_V$			 & [4, 6] mag		   & 5.2$\pm^{0.5}_{0.6}$ mag	& 5.7 mag  	\\
\hline
R$_o$ 			& \dots          & 3$\pm^{4}_{1}$ pc                    & 2 pc          \\          
L$_{bol}^{AGN}/10^{43}$ & \dots          & 6.0$\pm^{2.2}_{1.4}~erg~s^{-1}$ & 5.2$~erg~s^{-1}$   \\
N$_H^{torus}$/10$^{23}$   & \dots          & 7.0$\pm^{1.4}_{1.2}~cm^{-2}$    & 7.1$~cm^{-2}$       \\     
$M_{\rm torus}/10^{5}$  & \dots          & 2.2$\pm^{4.5}_{1.0}\,M_\odot$   & 1.7$\,M_\odot$     \\
\hline      
\end{tabular}						 
\caption{Clumpy model parameters, intervals considered as uniform priors, median and MAP values of the posteriors resulting from the fit of Mrk\,1066 nuclear SED.
Parameters: width of clouds angular distribution ($\sigma$), radial extent of the torus ($Y$), number of clouds along equatorial ray ($N_0$), index of the radial density profile ($q$),
inclination angle of the torus ($i$), optical depth per single cloud ($\tau_{V}$) and foreground extinction ($A_V$).}  
\label{tab2}
\end{table}

We fitted the nuclear SED of Mrk\,1066 considering reprocessed torus emission and foreground extinction (A$_V$). This 
extinction is introduced as another prior in the fit, and it accounts for additional dust along the line-of-sight (LOS), 
unrelated to the torus. 
\citet{Riffel10} presented extinction maps obtained from their integral field NIR data (using the Pa$\beta$/Br$\gamma$ line ratio), and
reported values ranging from $\sim$4 to 5.6 mag in the innermost region of Mrk\,1066. Thus, we considered the prior 
A$_V$=[4,6] mag in our fit and the IR extinction curve of \citet{Chiar06}. 

The fitted SED includes an \emph{HST}/NICMOS 1.6 \micron~nuclear flux (0.51$\pm$0.16 mJy; \citealt{Quillen01}), 
the GTC/CC 8.7 \micron~nuclear flux (63$\pm$9 mJy) and the GTC/CC spectrum, scaled to the latter flux. We
resampled the spectrum to $\sim$40 points, following the same methodology as in \citet{Alonso13}, and 
we did not include the spectral regions containing PAH emission bands in the fit. Finally, we used the 
25 \micron~\emph{Spitzer}/IRS flux reported in \citet{Sargsyan11} as an upper limit (2001 mJy) to constrain the 
longest MIR wavelengths. 

\begin{figure}
\includegraphics[width=9cm]{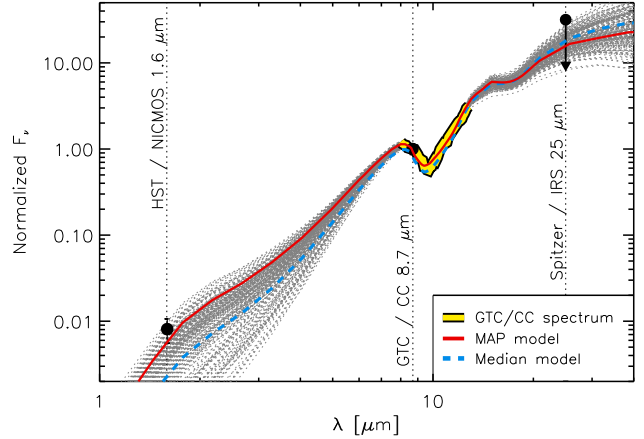}
\caption{High spatial resolution IR SED of Mrk\,1066 (thick yellow line: GTC/CC nuclear spectrum; black dots: \emph{HST}/NICMOS 
1.6 \micron~flux from 
\citealt{Quillen01}, GTC/CC 8.7 \micron~flux from this work and \emph{Spitzer}/IRS 25 \micron~flux from \citealt{Sargsyan11}), 
normalized to the 8.7 \micron~point. Solid red and dashed blue 
lines correspond to the MAP and median models respectively. Grey curves are the clumpy models sampled from the posterior and 
compatible with the data.
\label{fig5}}
\end{figure}

The result of the fitting process of the IR SED are the posterior 
distributions for the six parameters that describe the models (defined in Table \ref{tab2}), A$_V$ and the
vertical shift required to match the fluxes of a chosen model to an observed SED. 
These posteriors are shown in Figure \ref{A1} in Appendix \ref{appendixA}.
We can also translate the results into corresponding spectra, as shown in Figure \ref{fig5}. The red solid line
corresponds to the best-fitting model, described by the combination of parameters that maximizes the posterior 
(maximum-a-posteriori, MAP). The blue dashed line represents the model computed with the median value of each
parameter posterior. Finally, the grey dotted lines are all the clumpy SEDs reconstructed from the model
parameters sampled from the posterior, i.e. all the models that are compatible with the observations. We refer the 
reader to \citet{Asensio09,Asensio13} for further details on the Bayesian formalism.

It is clear from Figure \ref{fig5} that the MAP model is the one that better reproduces 
the observed data. In the following we will refer to MAP values, although the medians, with their corresponding
errors, are also reported in Table \ref{tab2} and shown in Figure \ref{A1} in Appendix \ref{appendixA}. 

An estimation of the AGN bolometric luminosity can be obtained from the vertical shift applied 
to the models to fit the data, which we allow to vary freely (see \citealt{Ramos09,Ramos11b} and \citealt{Alonso11} for
details).  
Using this shift, we obtain L$_{bol}^{AGN}$ = 5.2$\times10^{43}~erg~s^{-1}$, a value that  
can be compared with the bolometric luminosity estimated from the 2-10 keV luminosity. 
\citet{Marinucci12} 
reported a bolometric luminosity L$_{bol}^{X-ray}$ = 1.3$\times$10$^{44}~erg~s^{-1}$, once converted to our cosmology\footnote{\citet{Marinucci12} 
used a correction factor of 70 for obtaining the intrinsic 2-10 keV luminosity, as Mrk\,1066 is a Compton-thick candidate.}.
This luminosity was obtained using a large aperture, which contains part of the extended X-ray emission detected by \emph{Chandra} (see
Section \ref{discussion1}). Using the \emph{Chandra} data described in Section \ref{chandra},
we extracted a nuclear X-ray spectrum of the central 0.85\arcsec, and estimated a 2-10 keV luminosity of 1.5$\times10^{40}~erg~s^{-1}$.
Using the same correction factors as in \citet{Marinucci12}, we obtain L$_{bol}^{X-ray}$ = 2.0$\times10^{43}~erg~s^{-1}$.
Therefore, L$_{bol}^{AGN}$ derived from the fitted clumpy torus model is intermediate between 
the value reported by \citet{Marinucci12} and the estimation from \emph{Chandra} data calculated in this work. 

Using the MAP value of the optical extinction produced by the torus along the LOS (A$_V^{torus}$ = 375 mag), 
we can derive the column density of the obscuring material using the Galactic dust-to-gas ratio 
(N$_H^{torus}=1.9\times 10^{21}\times A_V^{torus}$; \citealt{Bohlin78}). This gives 
N$_H^{torus}=7.1\times10^{23}~cm^{-2}$ (see Table \ref{tab2}), which is compatible with X-ray observations showing that
Mrk\,1066 is a Compton-thick candidate \citep{Shu07}.


We can also estimate the torus gas mass from the fit, which in turn is a function of 
$\sigma$, $N_0$, $\tau_V$, $R_{\rm sub}$ and $Y$ (see Section~6.1 in \citealt{Nenkova08b}).
Using the MAP values reported in Table \ref{tab2}, we estimate 
$M_{\rm torus} = 1.7 \times 10^{5}\,M_\odot$. This is the gas mass in a clumpy torus of 
$\sim$2 pc radius, which is orders of magnitude smaller than the sizes probed by CO observations of Mrk\,1066 obtained with 
the 30 m single-dish IRAM telescope, which correspond to 24 kpc (M$_{H_2} = 2.7\times 10^{9}\,M_\odot$; \citealt{Kandalyan03}). 
\citet{Riffel11} also estimated a gas
mass of M$_{H_2} = 3.6\times 10^{7}\,M_\odot$ for the $\sim$70 pc radius circumnuclear disk of Mrk\,1066 
using the Gemini/NIFS data. This value
is still much larger than the torus mass estimated from our modelling, but again, we are talking about scales that differ
almost two orders of magnitude. 
As a comparison, we can look at the gas masses reported by \citet{Hicks09} for the inner parsecs of a sample of nearby Seyfert 
galaxies obtained from VLT/SINFONI measurements of the H$_2$1-0S(1) line at 2.1 \micron. In the case of the Circinus galaxy,
another Sy2 but at 4 Mpc distance only, they reported M$_{H_2} = 1.9\times 10^{6}\,M_\odot$ in a radius of 9 pc. 
This value is more comparable to Mrk\,1066's, probing similar spatial scales ($\sim$2 pc radius). 
More recently, \citet{Burillo14} reported a gas mass of $1.2 \times 10^{5}\,M_\odot$ for NGC\,1068 
(10 pc radius), using new CO(6-5) band 9 observations from the Atacama Large Millimeter/submillimeter Array (ALMA).


It is noteworthy the low torus inclination with respect to our LOS ($i$=17\degr). If we look at figure 15 
in \citet{Riffel11}, which 
shows a scheme of the inner $\sim$600 pc of Mrk\,1066, we can see that
the ionization cones are not exactly in the plane of the sky, which immediately 
implies a torus inclination different than edge-on ($i$=90\degr). The NW cone is indeed above the plane of the galaxy, 
whilst the SE cone is underneath. This picture is consistent with the kinematic modelling presented in \citet{Fischer13}. 
They reported a small opening angle for the ionization cones ($\theta$=25\degr) that agrees with the estimation
from \emph{HST}/WFPC observations of Mrk\,1066 \citep{Bower95} and with the large torus width derived from our
modelling ($\sigma$=68\degr).
However, \citet{Fischer13} reported a nearly edge-on torus inclination angle ($i$=80\degr), very different
to ours. We repeated the fit with clumpy torus models forcing $i$ to vary between 60\degr and 90\degr, but the results 
do not reproduce the \emph{HST}/NICMOS data point (the fitted models underestimate the NIR flux). In either way, 
the MAP torus model shown in Figure \ref{fig5} and the one resulting from the more restrictive $i$ prior only differ
9--12\% between 6 and 30 \micron, which is the maximum wavelength range considered in the following Sections.


\subsubsection{AGN+SB spectral decomposition fits}
\label{AGN_SB}

Taking advantage of the nuclear SED fitting with clumpy torus models performed in Section \ref{clumpy}, 
we can estimate the AGN contribution to the MIR emission of Mrk\,1066 on the spatial scales probed by 
the \emph{Spitzer}/IRS 5--38 \micron~and the GTC/CC 7.5--13 \micron~spectra ($\sim$830 and $\sim$60 pc respectively).

\begin{figure*}
{\par\includegraphics[width=8.6cm]{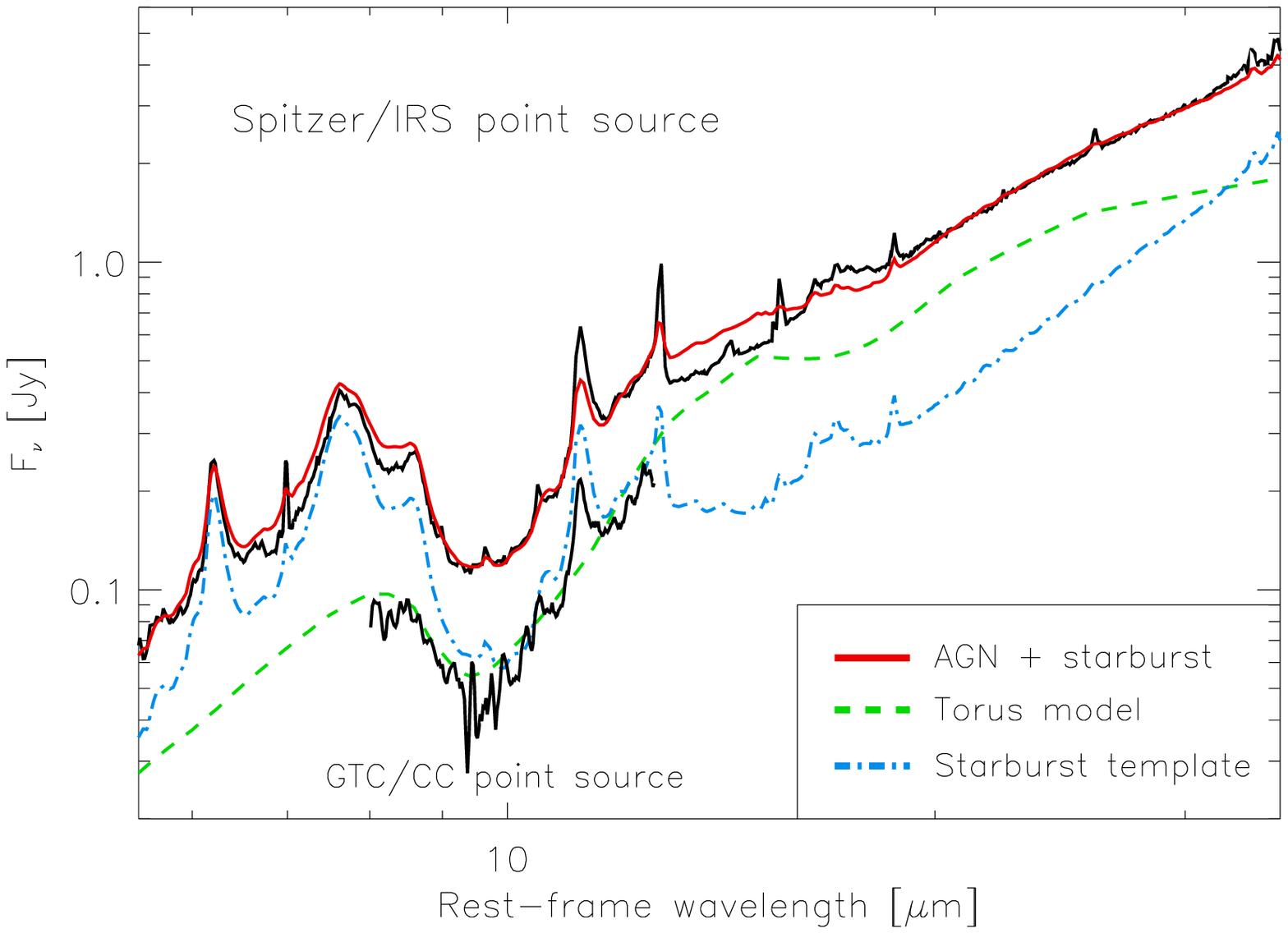}
\includegraphics[width=8.6cm]{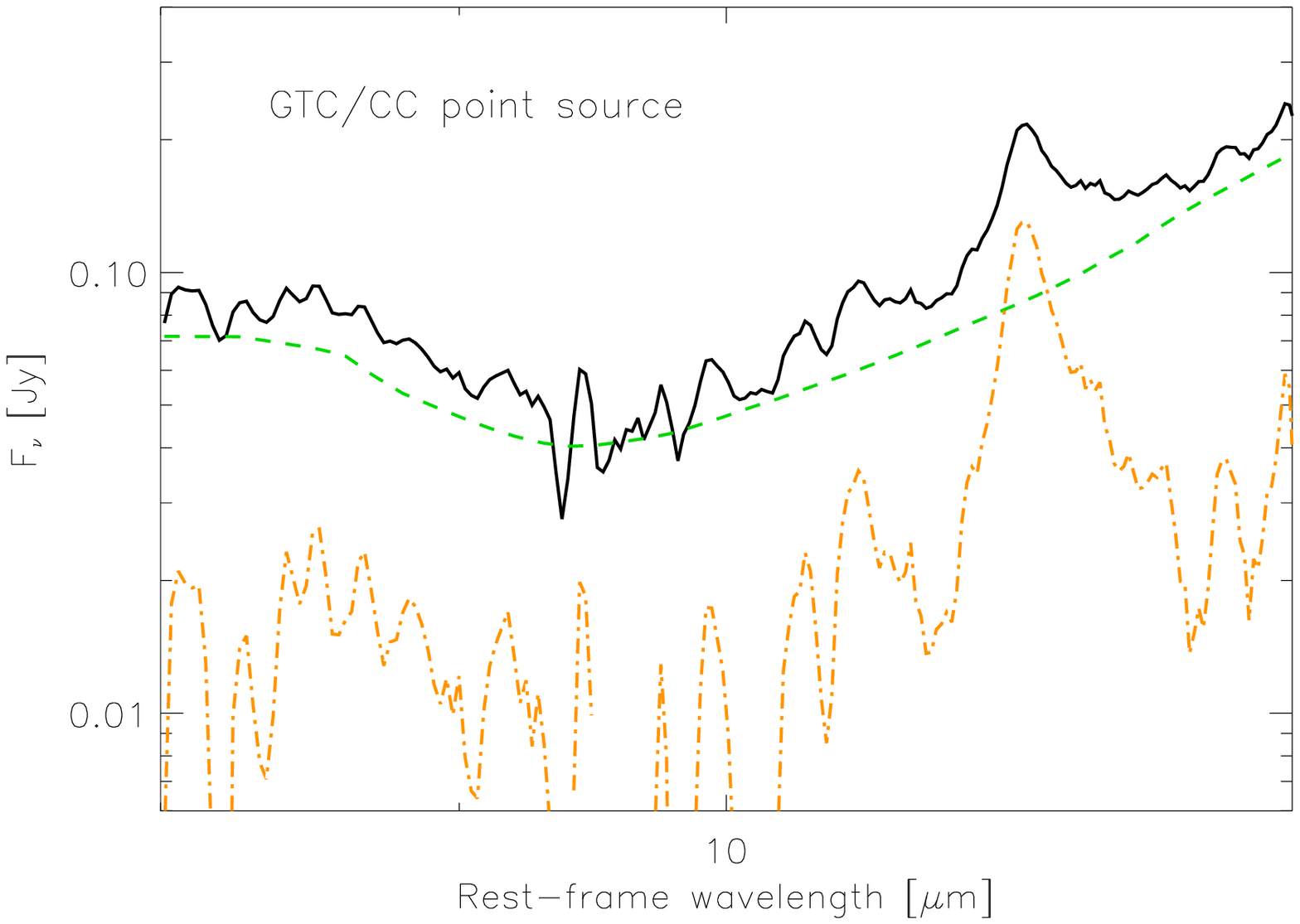}\par}
\caption{Left: Spectral decomposition of the \emph{Spitzer}/IRS spectrum extracted as point source. Black lines are the rest-frame 
\emph{Spitzer}/IRS and GTC/CC spectra. Green dashed and blue dot-dashed lines are the MAP torus model and the chosen starburst 
template respectively. Red solid line is the sum of the fitted starburst and AGN component. Right: GTC/CC spectrum and 
MAP torus model (solid black and green dashed lines respectively). Orange dot-dashed line corresponds to the 
residual from subtracting the MAP torus model from the GTC/CC nuclear spectrum.
\label{fig6}}
\end{figure*}

We took a simple approach to decompose the \emph{Spitzer}/IRS spectrum shown in Figure \ref{fig3} into AGN 
and starburst components. As the AGN template, we used the MAP clumpy torus 
model shown as a red solid line in Figure \ref{fig5}. The star-forming galaxy templates include
the average spectrum of local starbursts of \citet{Brandl06} and the templates of LIRGs in 
the range log($L_{IR}/L_{\odot}$) = 10.5--12 from \citet{Rieke09}. The fitting procedure is described in detail
in \citet{Alonso12b}, although they used an iterative method to perform the spectral decomposition, as they did not
have the privileged information from high angular resolution data that we do have for Mrk\,1066. In 
our case, we simply tried different combinations of the MAP torus model and the starburst 
templates, allowing for rescaling of the two components. We finally chose the starburst template that, together with the 
AGN component, minimized $\chi^2$, which corresponds to the log($L_{IR}/L_{\odot}$) = 11.25 LIRG template from \citet{Rieke09}.
The result of the fit for the \emph{Spitzer}/IRS spectrum, with a spatial resolution 
of 3.7\arcsec~(the SL slit width), is shown in the left panel of Figure \ref{fig6}.

The fit was done in the 6--30 \micron~range to avoid the edges and the slightly 
decreased signal-to-noise of the longest wavelengths. From Figure \ref{fig6} we see that, whilst
the starburst component dominates at shorter wavelengths ($\lambda<$15 \micron), with 30\% of AGN contribution at 
6 \micron, the AGN becomes dominant for $\lambda\ga$15 \micron, reaching 75\% at 15--25 \micron. At longer wavelengths, the AGN 
contribution smoothly decreases. This is consistent with the results found by \citet{Mullaney11} for a sample
of intermediate-luminosity AGN (L$_{2-10 keV} \sim 10^{42-44} erg~s^{-1}$), whose 6--100 \micron~SEDs are best described
by a broken power-law (representative of the AGN emission) that generally peaks between 15 and 20 \micron, 
and falls steeply at $\ga$40 \micron. In the case of Mrk\,1066, the starburst component becomes dominant again
at $\lambda>$30 \micron~(see left panel of Figure \ref{fig6}). 

The AGN contribution that we measured for Mrk\,1066 at 15--25 \micron, in the scales probed by \emph{Spitzer}/IRS (3.7\arcsec), is among
the largest reported by \citet{Alonso12b} for a volume-limited complete sample of 53 nearby LIRGs, which is representative of the
local LIRG population. 
Mrk\,1066 is nearly a LIRG with a Seyfert nucleus, 
if we consider the IR luminosity reported by \citet{Sargsyan11}, once converted to our cosmology, log(L$_{IR}$/L$_{\odot}$)=10.9. 
Using this value, we can estimate the AGN bolometric contribution to the IR luminosity, L$_{bol}^{X-ray}$/L$_{IR}$=0.43\footnote{ 
Here we use the value of L$_{bol}^{X-ray}$ from \citet{Marinucci12}, which is consistent with the luminosities used in
\citet{Alonso12b}. Note that in the case of Mrk\,1066, this L$_{bol}^{X-ray}$ constitutes 
an upper limit to the real L$_{bol}^{AGN}$ (see Section \ref{clumpy}).}. 
Therefore, Mrk\,1066 belongs to the 8\% of local LIRGs studied in \citet{Alonso12b} that have a significant AGN 
bolometric contribution to the IR luminosity (L$_{bol}^{X-ray}$/L$_{IR}>$0.25), which are indeed 
those classified as Seyferts. 

In the left panel of Figure \ref{fig6} we also show the GTC/CC 8--12.7 \micron~spectrum, extracted as point source, 
for comparison. 
Without applying any scaling, the GTC/CC spectrum coincides with the AGN component obtained from the \emph{Spitzer}/IRS fit.
This is expected, as, by definition, the AGN component always has to be the same. In the right panel of Figure \ref{fig6} we plot 
the GTC/CC nuclear spectrum and the MAP torus model in Figure \ref{fig5}. According to the fit performed in Section 
\ref{clumpy}, the AGN contribution on the scales probed by the GTC/CC spectrum ($\sim$60 pc) dominates the nuclear
MIR emission, varying between 90\% and 100\%, depending on the wavelength (we chose two featureless regions of the 
spectra, centred at 8.2 and 12 \micron). The dot-dashed line in the right panel of Figure \ref{fig6} corresponds to 
the residual from subtracting the MAP torus model from the GTC/CC nuclear spectrum. 


The contribution from star formation to 
the nuclear GTC/CC spectrum of Mrk\,1066 (i.e. the inner 0.26\arcsec$\times$0.52\arcsec~$\approx$ 60 pc$\times$120 pc) 
is mainly concentrated on the PAH 
features at 8.6 and 11.3 \micron\footnote{With the exception of the silicate feature, the clumpy torus models 
do not account for spectral features.}, as it can be seen from the right panel of Figure \ref{fig6}.
Whilst the contribution of star formation to the weaker 8.6 \micron~emission is only $\sim$20\%, 
the 11.3 \micron~band is mostly due to star formation ($\sim$60\%). In the case of the \emph{Spitzer}/IRS 3.7\arcsec~aperture, 
the starburst contribution to the two PAH features is $\sim$70\%.

Summarizing, for the case of Mrk\,1066, the AGN component dominates the MIR at $\lambda<$15 \micron~on scales of 
$\sim$60 pc (90--100\%), and decreases down to 35--50\% when the 8--12.5 \micron~emission of the central $\sim$830 pc is 
considered\footnote{Also measured at 8.2 and 12 \micron.}. Instead, if we look at longer wavelengths ($\lambda>$15 \micron) 
the AGN component probed by the \emph{Spitzer}/IRS 3.7\arcsec~aperture reaches 75\% at 15--25 \micron~(i.e. at the peak of AGN emission), 
which is among the highest percentages in the reference sample of LIRGs studied in \citet{Alonso12b}.

\subsubsection{MIR spectroscopy of the knots}
\label{extended_spectroscopy}

\begin{figure*}
\par{
\includegraphics[width=8.6cm]{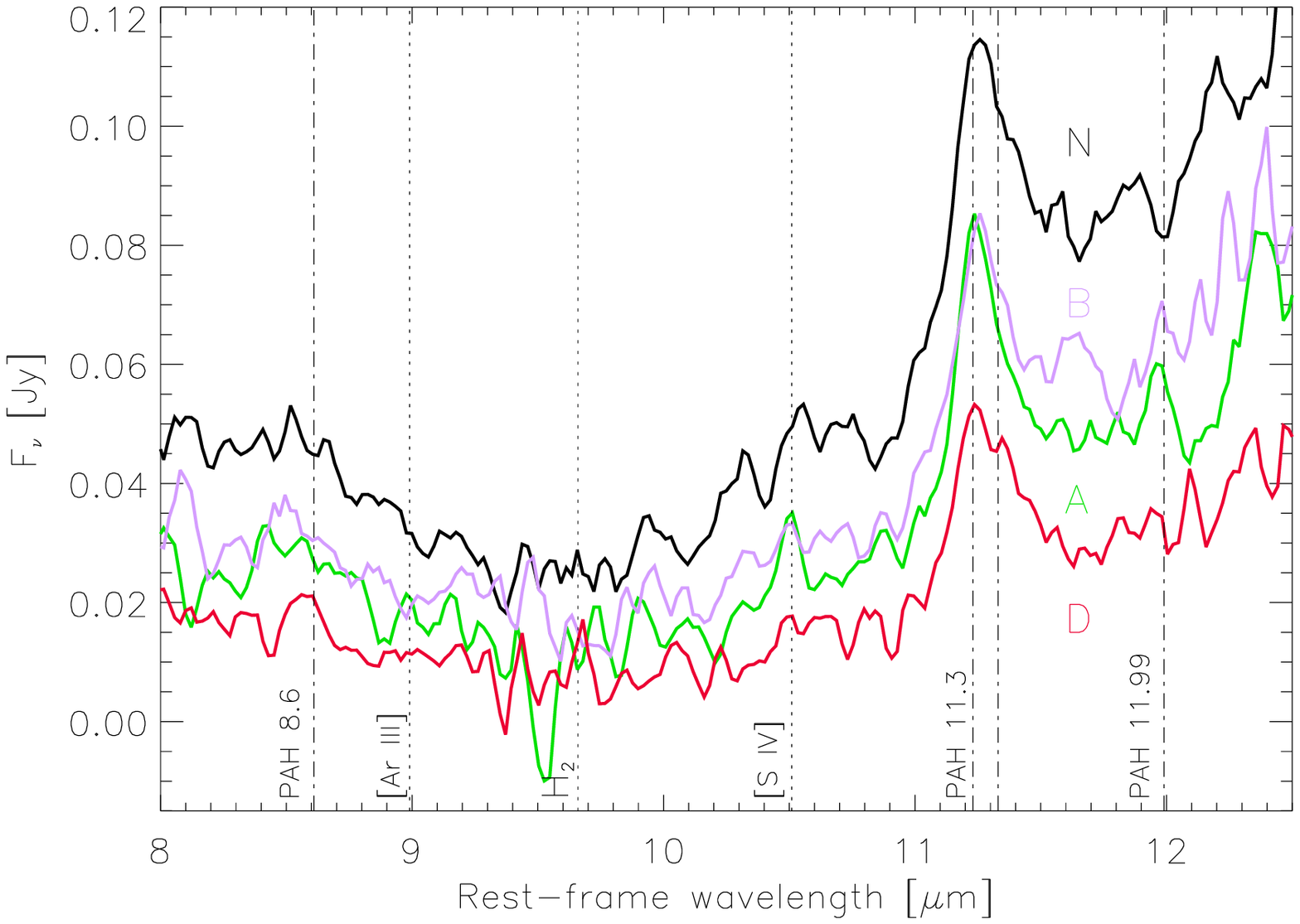}
\includegraphics[width=8.6cm]{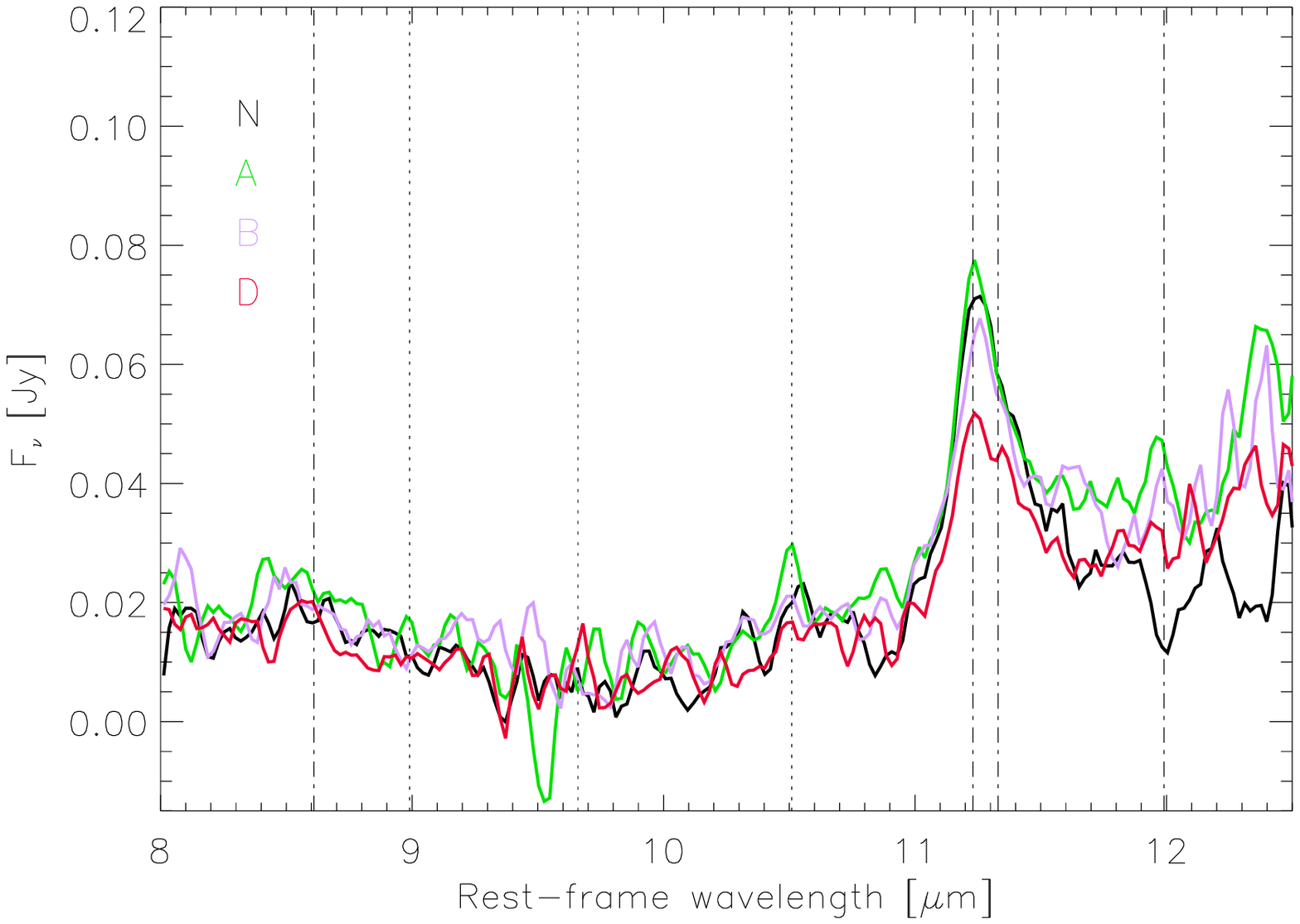}\par}
\caption{Left: GTC/CC N-band spectra of the nucleus (black line) and knots A, B and D (green, purple and red line respectively)
extracted as extended sources, using an aperture of 0.52\arcsec$\times$0.4\arcsec~and centered in the positions listed in Table \ref{tab1}.
Right: AGN-subtracted GTC/CC N-band spectra of the knots.
Vertical dotted lines indicate the position of the emission lines and the PAH features. 
\label{fig4}}
\end{figure*}

From the analysis of the GTC/CC MIR nuclear spectrum of Mrk\,1066 it is clear that there is star formation 
activity taking place in the inner $\sim$60 pc of the galaxy. Now we can take advantage 
of the spatial information afforded by the GTC/CC spectroscopy and extract spectra at the location of knots 
A, B and D and the nucleus. Thus, we extracted the four spectra as extended sources, centered in the positions listed in Table \ref{tab1} 
and using apertures of 0.52\arcsec$\times$0.4\arcsec. This is the maximum aperture that we can use avoiding overlap between the different knots. 
These spectra are shown in the left panel of Figure \ref{fig4}. The nuclear spectrum (N)
is the brightest, whilst knot D is the dimmest ($\sim$1\arcsec~NW). 

Only in the case of knot D we resolve the double peak
of the 11.3 \micron~PAH feature. 
Besides, we marginally detect the [S IV]10.5 \micron~fine structure line in the 
spectrum of knot A. This line is thought to originate in the NLR, as it is well correlated with [Ne V] and [O IV],
and therefore it is widely used as AGN tracer (see \citealt{Dasyra11} and references therein). However, as shown by \citet{Pereira10}, 
this emission line can be also produced in star-forming regions, due to its low ionization potential (35 eV), 
similar to that of the [O III]$\lambda$5007 \AA~line \citep{Trouille11}. 

The spectra shown in the left panel of Figure \ref{fig4} have contributions from both AGN and star formation. 
Therefore, in order to minimize as much as possible the AGN contamination from the spectra of the knots, 
we estimated the AGN contribution at their positions using the GTC/CC 8.7 \micron~images, before and after 
PSF subtraction (see top panels of Figure \ref{fig2}).
 
\begin{figure*}
\includegraphics[width=12cm]{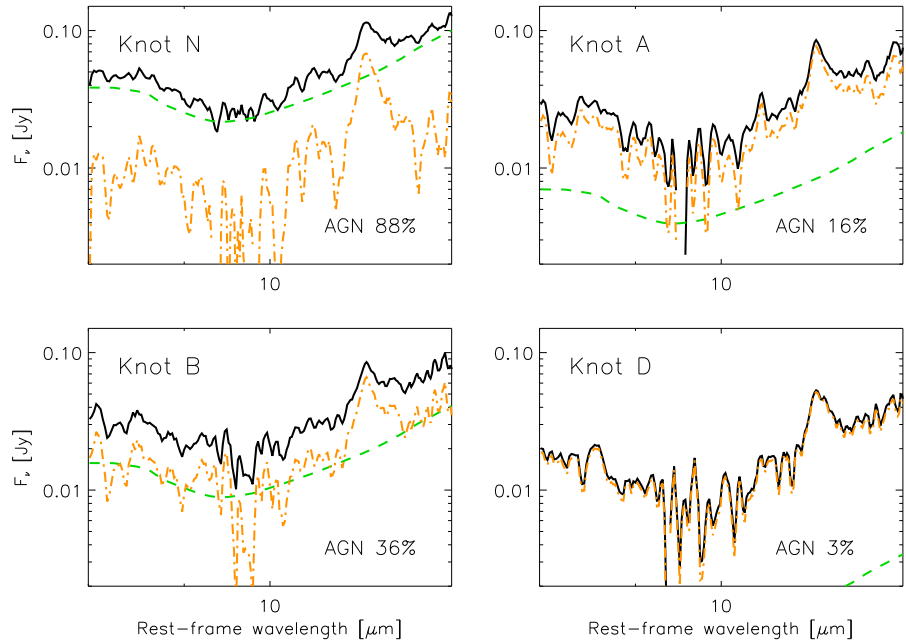}
\caption{Same as the right panel of Figure \ref{fig6}, but for the GTC/CC MIR spectra of the knots, extracted as extended sources.
Black lines are the rest-frame spectra, 
green dashed line is the MAP torus model, once scaled to the percentages indicated in each panel, which correspond to the 
corresponding AGN contributions. Dot-dashed orange lines are the results from the subtraction of the scaled AGN 
component (i.e. the green dashed line) from the spectrum of each knot.   
\label{fig7}}
\end{figure*}

We obtained 8.7 \micron~fluxes in the two images at the positions of the four knots, using the same aperture employed 
for extracting the spectra (0.4\arcsec). Then, first we compared the measurements before and after PSF-subtraction
and estimated the AGN contribution to the total fluxes in each of the knots. 
We find that the AGN represents 88\%, 16\%, 36\% and 3\% in knots N, A, B and D respectively. 
Second, we scaled the MAP torus model obtained from the 
fit of the GTC/CC nuclear spectrum (see Figure \ref{fig5}), to the spectrum of knot N, extracted as an extended 
source. To do that, we calculated the scale factor between the GTC/CC nuclear spectra (extracted as point and extended source)
and applied it to the MAP torus model. Finally, we multiplied the scaled MAP torus model by the percentages obtained from 
the comparison between the MIR images, and subtracted the AGN component from the spectra of the knots (see Figure \ref{fig7}). 



The residuals from the subtraction, i.e. the spectra of the knots without the AGN contribution, are 
also shown in the right panel of Figure \ref{fig4}. We note that using this method we are just subtracting
the unresolved MIR emission, dominated by dust within the torus, but not the extended NLR emission.

The AGN-subtracted spectra of the knots are flatter than those without subtraction (shown in the left panel
of Figure \ref{fig4}). This implies 
that the deep silicate feature observed in the spectrum of the nucleus is partly produced by both the 
obscuring dusty torus of Mrk\,1066 and the foreground extinction that we considered in the fit (A$_V$=4--6 mag).

We measured the flux and equivalent width (EW) of the 11.3 \micron~PAH feature in the GTC/CC spectra of the knots
before and after subtracting the AGN contribution (shown in Figure \ref{fig4}). We followed the method described in 
\citet{Hernan11} and implemented for ground-based spectroscopy by \citet{Esquej14}. We fitted a local 
continuum using two narrow bands (10.75--11 \micron~and 11.65--11.9 \micron) adjacent to the 11.3 \micron~PAH 
feature, and we integrated the flux above the continuum in the rest-frame range 11.05--11.55 \micron. 
Uncertainties are estimated by performing Monte Carlo simulations, as described in detail in \citet{Esquej14}.
In the case of the EWs measured from the AGN-subtracted spectra, the errors also include the uncertainties
in the fluxes calculated from the GTC/CC 8.7 \micron~image and the PSF-subtracted 8.7 \micron~map. 



\begin{table}
\centering
\footnotesize
\begin{tabular}{ccccc}
\hline
\hline
ID & L$_{11.3 \micron}$ & SFR & EW & AGN-sub EW  \\
& ($\times10^{41}$ erg~s$^{-1}$) & (M$_\odot$~yr$^{-1}$) & (\micron) & (\micron) \\
\hline
N     &  1.31$\pm$0.07 & 0.32$\pm$0.02 & 0.34$\pm$0.02  & 1.89$\pm$0.32   		\\ 
A     &  1.00$\pm$0.06 & 0.25$\pm$0.01 & 0.46$\pm$0.03  & 0.63$\pm^{0.42}_{0.17}$	\\ 
B     &  0.99$\pm$0.06 & 0.25$\pm$0.01 & 0.39$\pm$0.02  & 0.80$\pm^{0.45}_{0.41}$	\\  
D     &  0.78$\pm$0.05 & 0.19$\pm$0.01 & 0.61$\pm$0.05  & 0.67$\pm^{1.03}_{0.06}$	\\
\hline
\end{tabular}
\caption{Measurements of the 11.3 \micron~PAH feature in the knots before and after subtracting the AGN contribution. 
The values of L$_{11.3 \micron}$ and SFR before and after AGN subtraction are compatible within the errors. 
Errors in the last column include the uncertainties associated with AGN subtraction.}
\label{tab3}
\end{table}

\begin{figure}
\includegraphics[width=8.6cm]{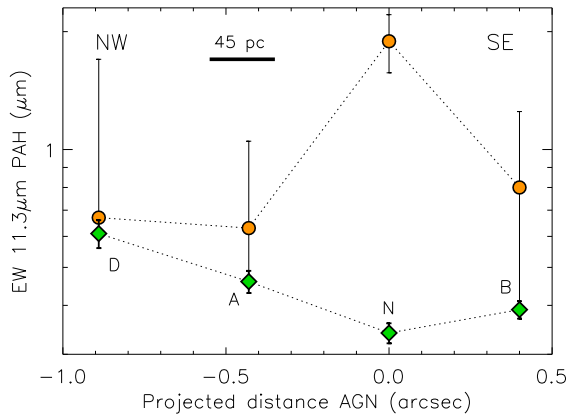}
\caption{Spatial variation of the 11.3 \micron~PAH feature EW before and after AGN subtraction (green diamonds and orange circles, 
respectively).    
\label{fig11}}
\end{figure}

Before subtracting the AGN component from the spectra of the knots, the EW of the 11.3 \micron~PAH feature is lower in the 
nucleus than in the knots (see Table \ref{tab3} and Figure \ref{fig11}). 
Knot D, which is the most distant from the AGN, shows 
the largest value of the EW. These values are equal or larger than those
reported by \citet{Esquej14} using the same technique and probing similar scales (e.g. EW$_{11.3}$=
0.36$\pm$0.02 \micron~for the star-forming Seyfert galaxy NGC\,1808).  

If we look at the PAH EWs after AGN subtraction (see Table \ref{tab3} and Figure \ref{fig11}), we find that the values
increase, specially in the nucleus (1.9$\pm$0.3 \micron). The lowest EW corresponds to knot A (0.6$\pm^{0.4}_{0.2}$ \micron), 
and the EW of knot D remains the same within the errors. The nuclear EW is similar to the values reported in figure 6 of 
\citet{Alonso14} for regions at $\sim$200--300 pc SE and NW of the nucleus (1.4--1.5$\pm$0.3 \micron)\footnote{We note that \citet{Alonso14}  
did not subtract the AGN component from the GTC/CC spectra.}. The EWs measured
for knots A, B and D, on the other hand, are lower. 

The AGN-corrected EWs are consistent with those measured for pure starbursts 
in \citet{Hernan11} and for extra-nuclear star-forming regions in nearby LIRGs \citep{Diaz10}. 
Thus, here we propose a reliable method to subtract the AGN continuum from MIR spectra, allowing 
to derive EWs which are representative of star formation. 


Finally, using the 11.3 \micron~PAH luminosities measured for the knots before and after subtracting the AGN
contribution from the spectra, we can calculate SFRs 
following the empirical relation derived in \citet{Diamond12}: 

\[
SFR(M_{\odot}~yr^{-1}) = 9.6\times10^{-9} L(11.3~\micron; L_\odot).
\]

The 11.3 \micron~PAH luminosities and corresponding SFRs are reported in Table \ref{tab3}. 
Before and after AGN subtraction, we measured SFRs ranging between 0.2 and 0.3 M$_{\odot}~yr^{-1}$, which are among the 
largest values measured by \citet{Esquej14} for a sample of local Seyfert galaxies in regions of $\sim$65 pc in 
size\footnote{Note that neither the luminosities nor the SFRs reported in Table \ref{tab3} have been multiplied by
the factor of two derived by \citet{Smith07} and used in \citet{Esquej14}.}. 




\section{Discussion}
\label{discussion}

\subsection{The origin of the circumnuclear emission of Mrk\,1066}
\label{discussion1}


From the similarity between the MIR images (with and without PSF subtraction) and
the Pa$\beta$ and Br$\gamma$ images presented in Section \ref{imaging} (see Figure \ref{fig2}), we can confirm that the GTC/CC Si-2 filter is 
tracing both emission from the NLR and star formation\footnote{The Si-2 filter includes the contribution from the 8.6 
\micron~PAH feature and its underlying continuum.}, and this emission consists on four discrete 
knots and an extended component. The coincidence between the extended 8.7 \micron~emission and hydrogen recombination
lines is common in IR bright galaxies \citep{Helou04,Alonso06,Diaz08} and it implies 
that the 8.7 \micron~emission is due to star formation and/or dust in the NLR in the case of 
AGN \citep{Radomski03,Packham05}. In the case of Mrk\,1066, part of the extended MIR emission could also 
come from the oval structure of $\sim$350 pc radius detected from the emission line kinematics \citep{Riffel11}.

Based on the comparison between the soft X-ray and MIR morphologies of Mrk\,1066 (see Figure \ref{fig8}), 
it seems clear that the soft X-ray emission is not tracing the 
NLR of this Sy2 galaxy, in contradiction with the most accepted interpretation for its origin. It is generally assumed 
that the soft X-ray emission 
of Sy2s is produced by gas photoionized by the nuclear continuum \citep{Kinkhabwala02,Ogle03,Bianchi06,Guainazzi07}. 
Using a small sample of eight Sy2 galaxies, \citet{Bianchi06} compared \emph{Chandra} soft X-ray images with narrow-band optical 
images obtained with the \emph{HST}, containing the [O III]$\lambda$5007 \AA~emission\footnote{[O III]
is generally used to trace the NLR emission in AGN.}. They found a good coincidence 
between the soft X-ray and [O III] morphologies, on scales of hundreds of parsecs. However, we find that neither the 
MIR not the NIR morphologies, which also trace the NLR emission, as well as star formation, match the soft X-ray 
emission of Mrk\,1066. 

\begin{figure*}
{\par\includegraphics[width=5.9cm]{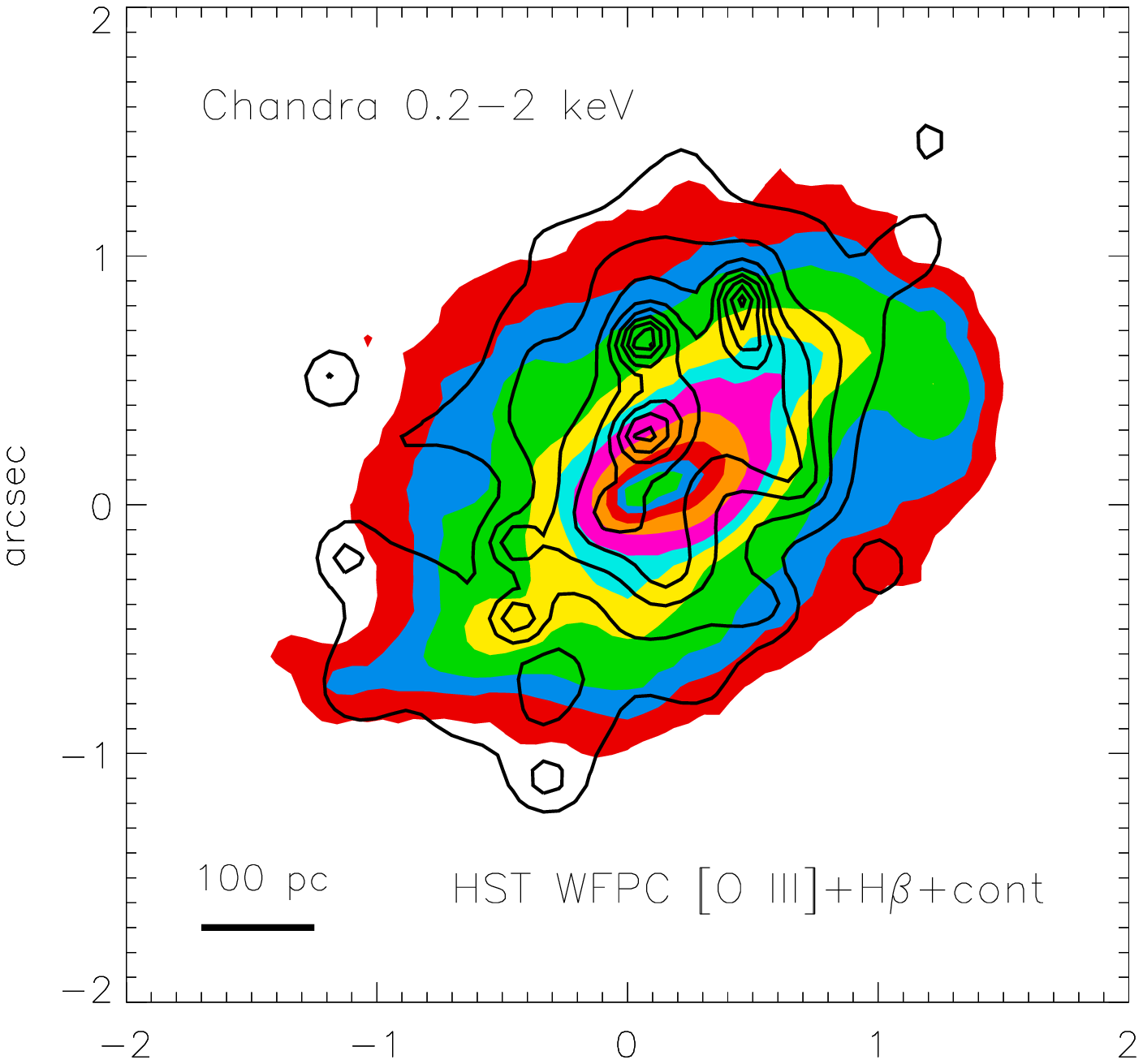}
\includegraphics[width=5.9cm]{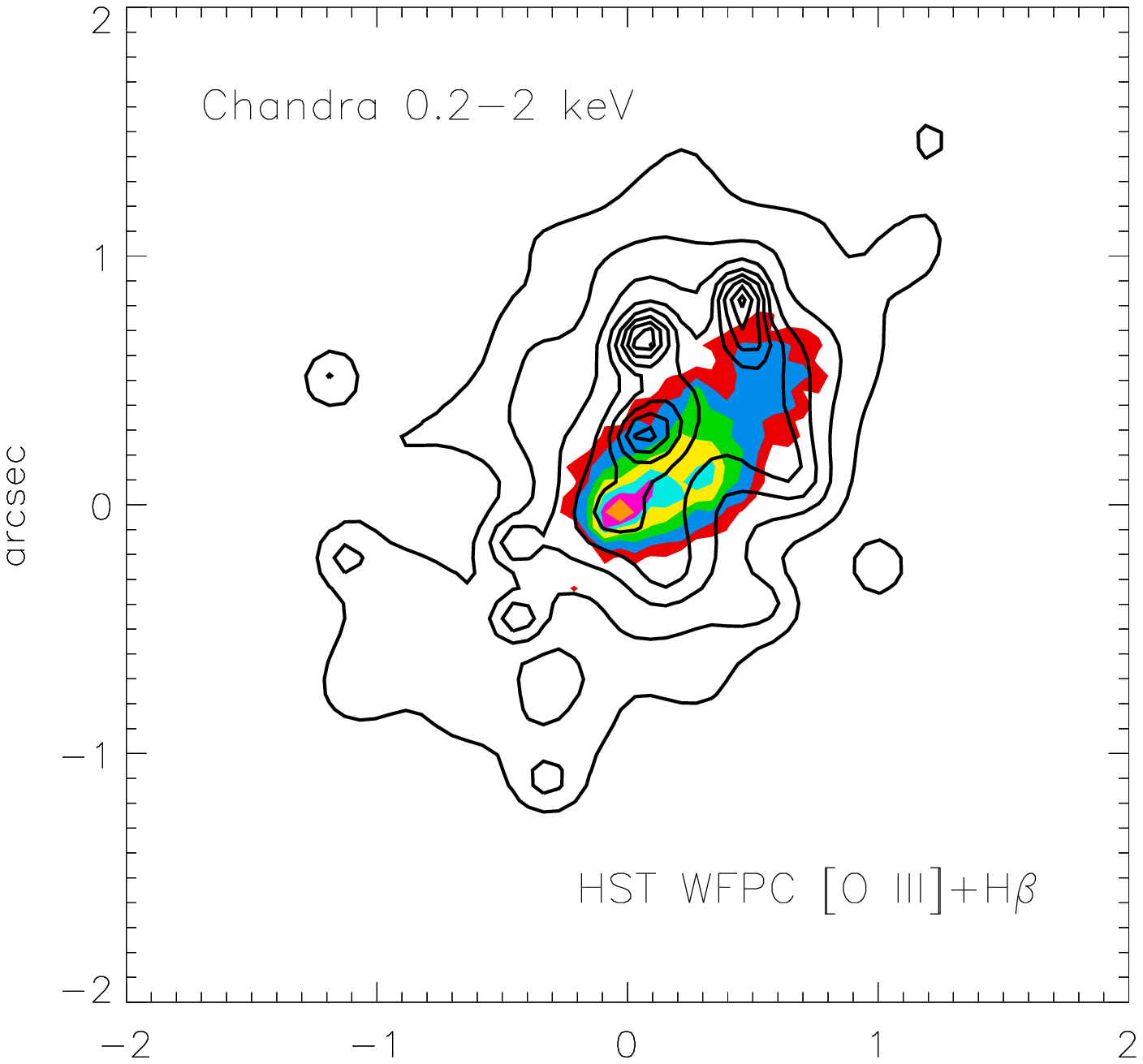}
\includegraphics[width=5.9cm]{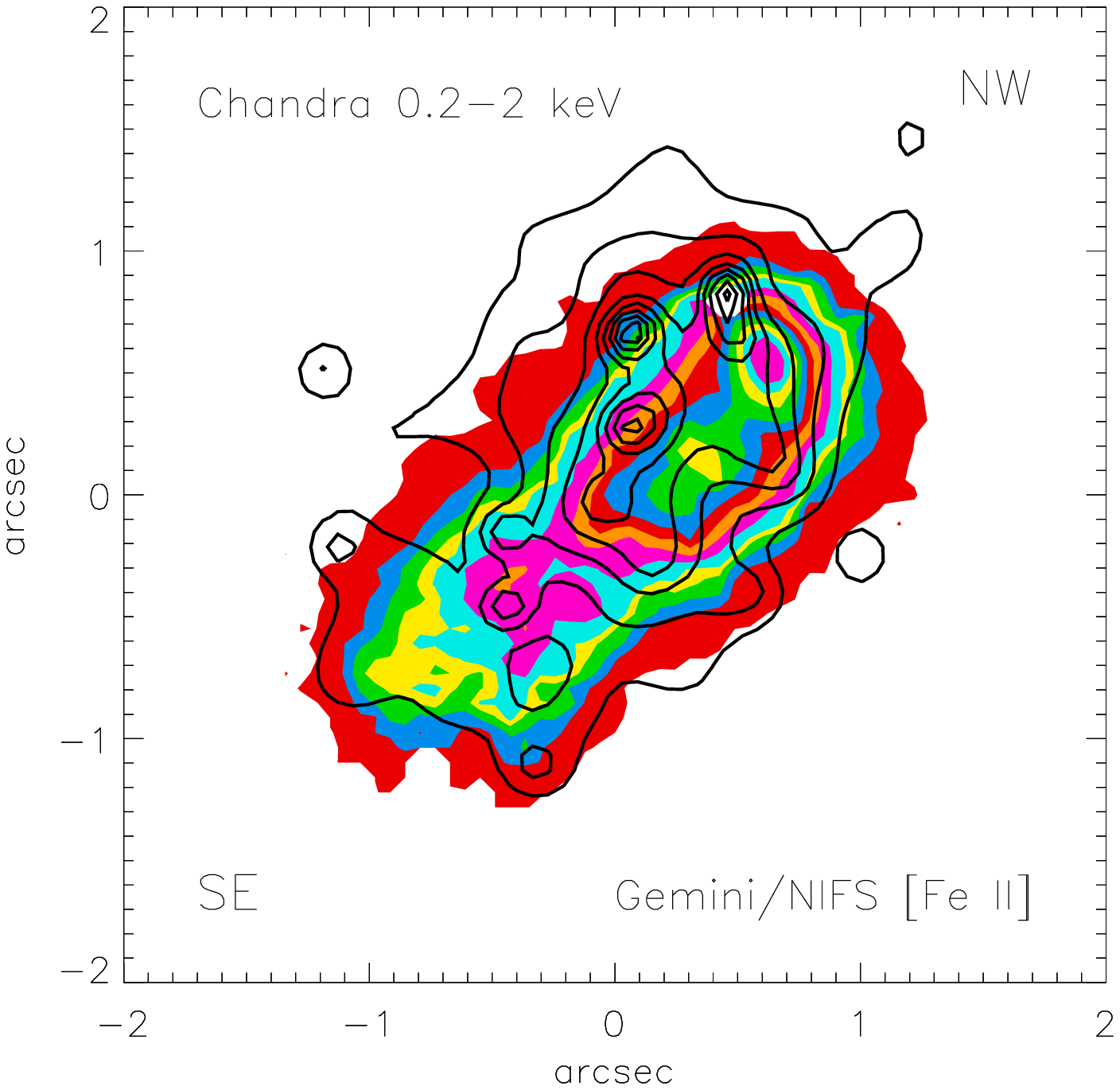}
\includegraphics[width=5.9cm]{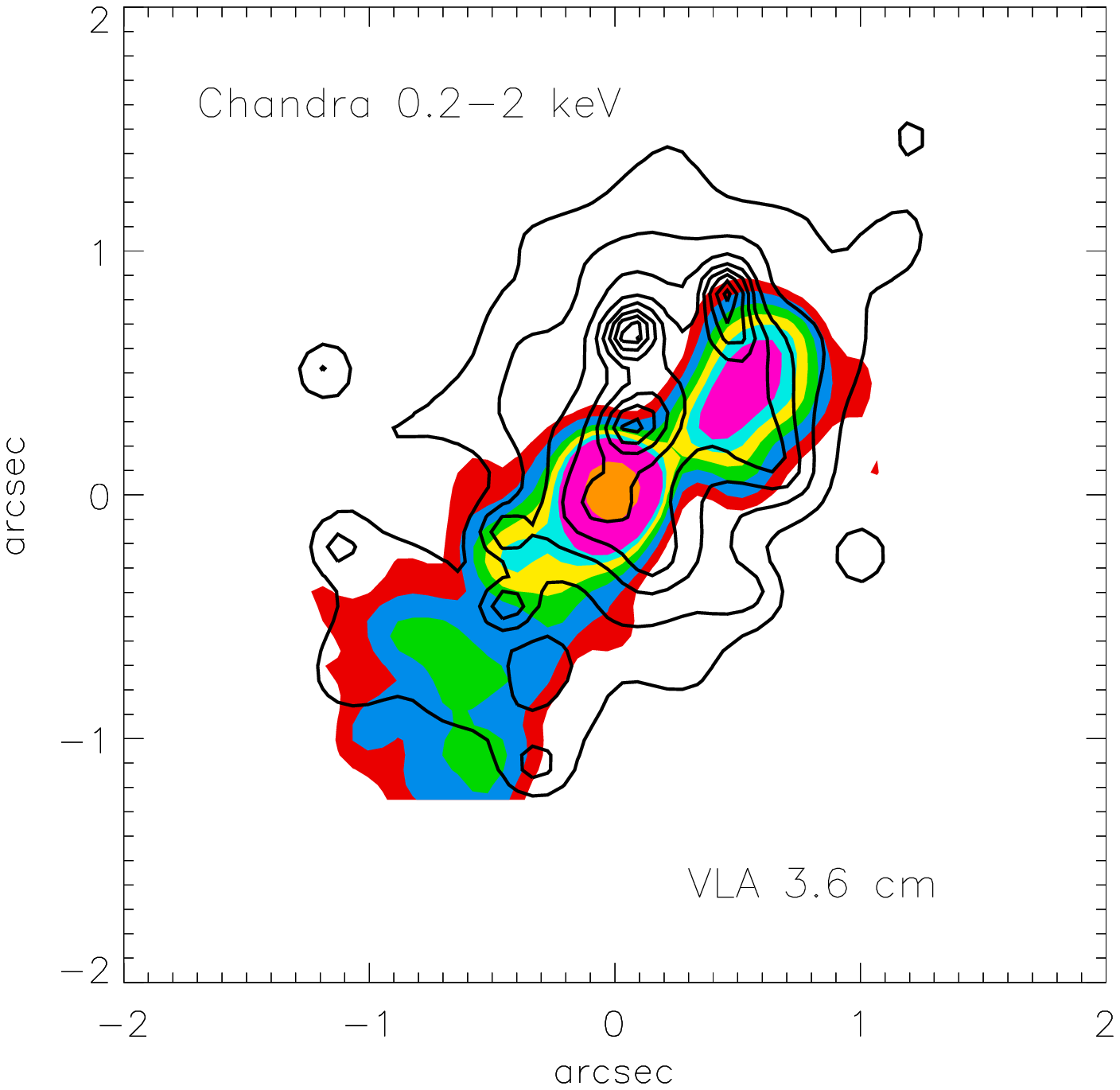}\par}
\caption{Contours of the soft (0.5--2 keV) X-ray emission of Mrk\,1066 from \emph{Chandra}/ACIS imaging data (black lines), 
overlaid on the \emph{HST}/WFPC [O III], continuum-subtracted [O III], Gemini/NIFS [Fe II]$\lambda$1.257 \micron~and 
VLA 3.6 cm contours (in colour).
The WFPC, NIFS and VLA images have been interpolated to the pixel size of the \emph{Chandra}/ACIS image (0.06\arcsec). 
\label{fig9}}
\end{figure*}

In the top left panel of Figure \ref{fig9} we compare the contours of the soft X-ray emission (in black) with those of 
the HST/WFPC [O III]+H$\beta$+continuum \AA~image (described in Section \ref{hst}). Overall, the soft X-ray emission 
agrees relatively well with the [O III] emission. However, once we subtract the continuum emission, by using the 
image taken in the adjancent filter F547N (see top right panel of Figure \ref{fig9}), we find a completely different 
[O III] morphology, that does not match the soft X-ray emission. Therefore, it would be necessary to compare  
the continuum-subtracted [O III] and soft X-ray morphologies of a large sample of galaxies to confirm/discard
a common source of ionization.

Alternatively, a thermal origin for the extended soft X-ray emission of star-forming active galaxies
as Mrk\,1066 has been also claimed (see e.g. \citealt{Levenson01}), 
and one third of Sy2 galaxies have been found to have thermal soft X-ray emission \citep{Turner97}.
In the bottom panels of Figure \ref{fig9} we compare 
the soft X-ray contours of Mrk\,1066 with the Gemini/NIFS [Fe II]$\lambda$1.257 \micron~and VLA 3.6 cm maps from \citet{Riffel10}, 
both showing their peaks of emission along the NW cone. 

Strong [Fe II] emission is common in Sy2 galaxies, but there are three mechanisms that can produce it, namely, 
1) AGN photoionization, 2) radio jet interactions with the surrounding medium and 3) fast shocks associated with supernova 
remnants in starburst regions (see \citealt{Ramos09} and references therein). In the case of Mrk\,1066, the peaks of 
the [Fe II] and radio emission perfectly match, as shown in Figure \ref{fig9} and in figure 4 of \citet{Riffel10}, indicating that the 
[Fe II] emission is at least partly produced by the radio jet interaction with the ISM.

From the bottom right panel of Figure \ref{fig9} we can see that the three soft X-ray knots are just outside 
the jet cocoon. \citet{Riffel11} claimed that the outflow of ionized gas to the NW would be
produced by the radio jet interaction, with the jet pushing away the ISM material and exciting it. This scenario would explain the 
increase in velocity dispersion shown in figure 5 of \citet{Riffel11}, which exactly coincides with the position of the most luminous 
soft X-ray knot. Moreover, high velocity [Fe II] gas is detected at the position of the central soft X-ray knot (see figure 4 
in \citealt{Riffel11}), also suggesting a thermal origin for both the [Fe II] and the extended soft X-ray emission.

\begin{figure}
\centering
{\par\includegraphics[width=5.8cm]{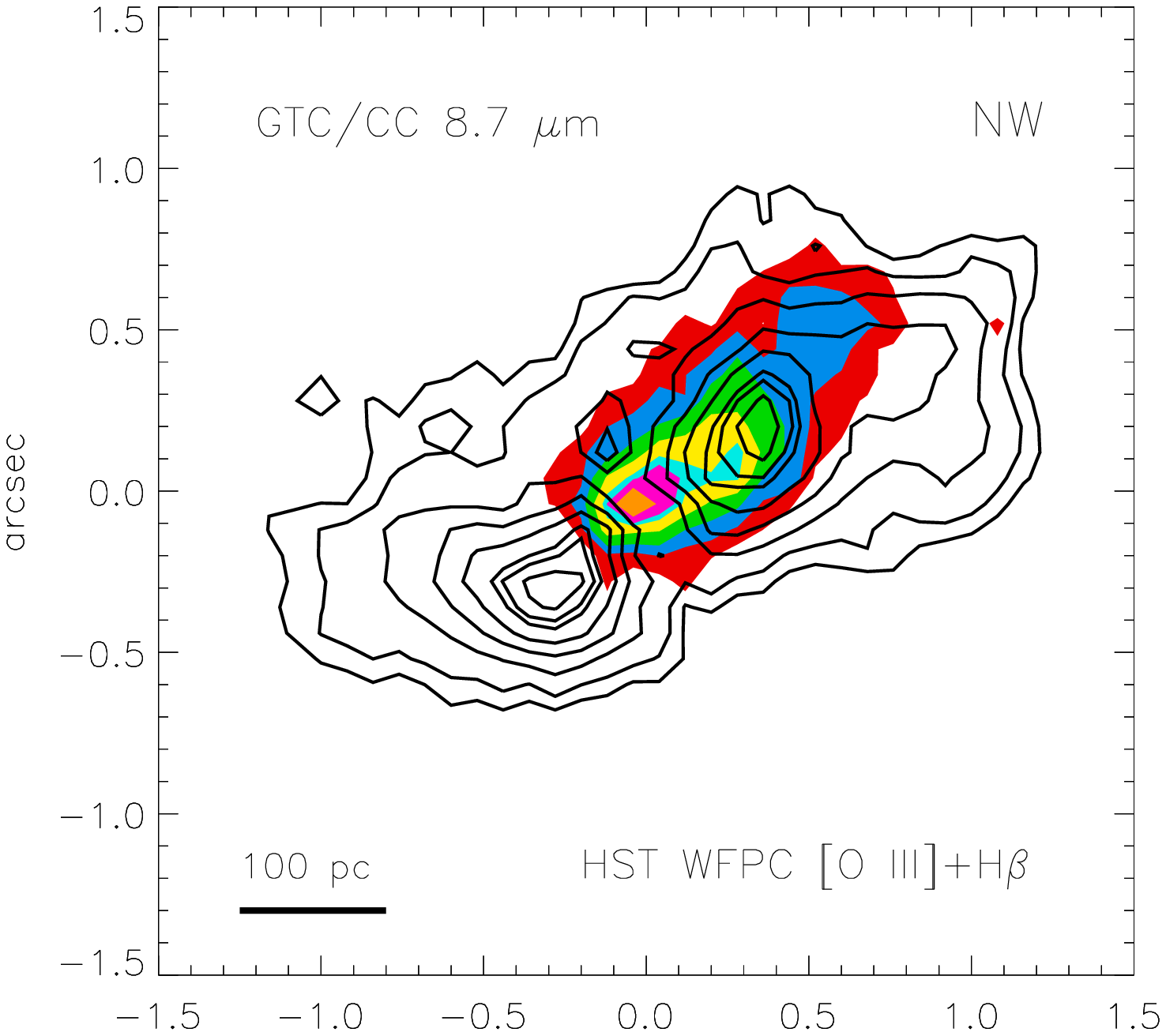}
\includegraphics[width=5.8cm]{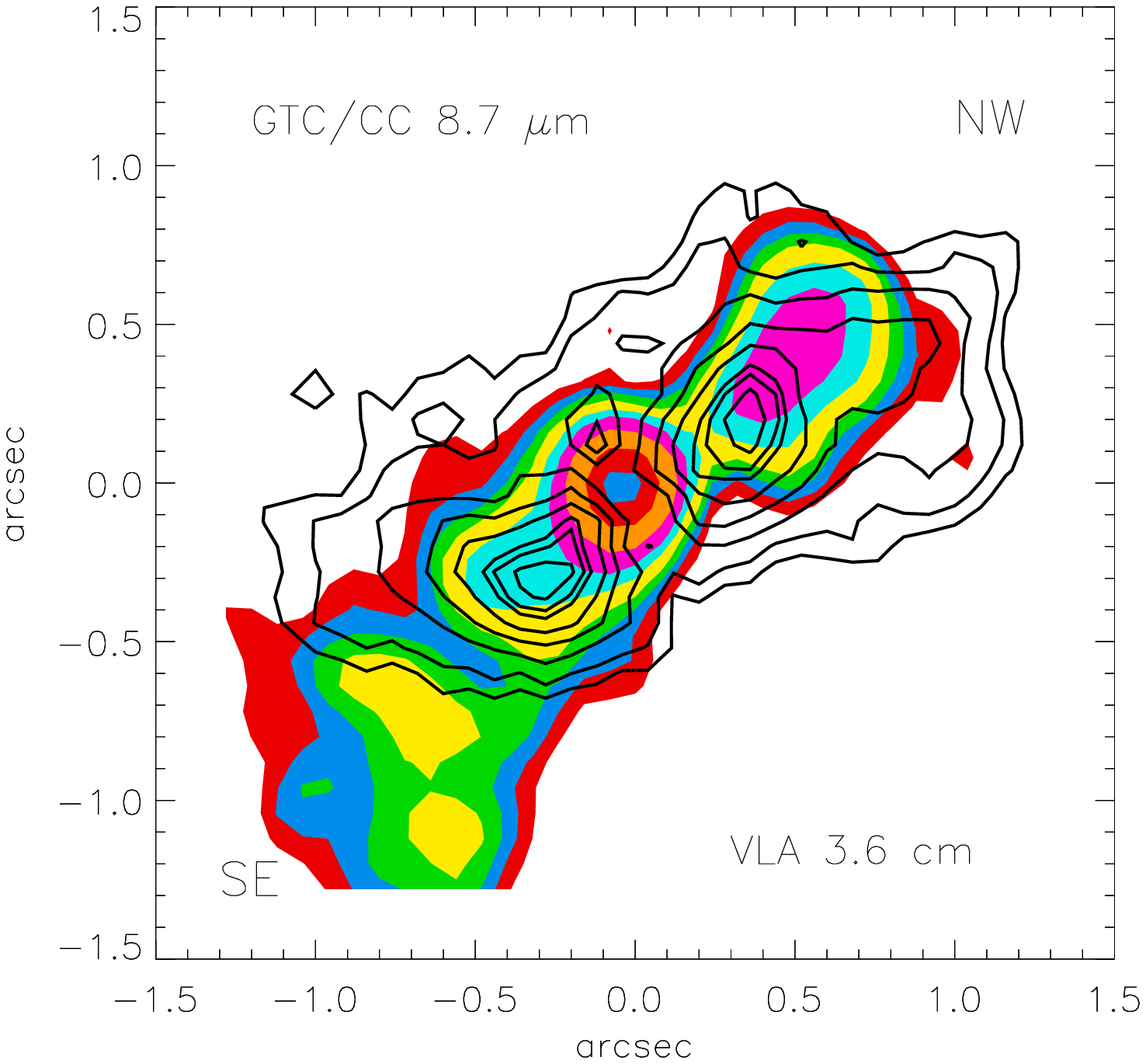}\par}
\caption{Contours of the PSF-subtracted GTC/CC 8.7 \micron~image (black lines) overlaid on the \emph{HST}/WFPC 
[O III] continuum-subtracted (top panel) and VLA 3.6 cm colour contours (bottom panel), 
interpolated to the pixel size of GTC/CC (0.08\arcsec). 
\label{fig10}}
\end{figure}

On the other hand, the NIR and MIR knots are within the cone and the jet, and the extended IR emission is cospatial with the 
NLR emission. In Figure \ref{fig10} we show the continuum-subtracted [O III] image and the 3.6 cm radio contours, 
interpolated to the pixel size 
of the GTC/CC image (0.08\arcsec), with the PSF-subtracted 8.7 \micron~contours overlaid. The [O III] emission is
elongated in the same direction as the IR emission. The line-emitting gas of the SE part of the NLR of Mrk\,1066 is detected 
in the NIR and MIR, but not in the optical, and this is likely due to extinction from the 
foreground galaxy. In fact, Figure \ref{fig10} completely agrees with the scheme of the central parsecs
of the galaxy shown in 
figure 15 of \citet{Riffel11}, in which the SE cone would be obscured by the rotating disk 
whose major axis coincides with the stellar disk. They claim that the bulk of [O III] emission would be produced in 
the biconical outflow oriented along the radio jet and ionization cones, as shown in in Figure \ref{fig10} (PA=315\degr). 
Besides, this scheme is also compatible 
with an intermediate orientation of the obscuring torus, which collimates the emission of the 
ionization cones (see Section \ref{clumpy}). 

We also note that the MIR emission is more extended to the NW than
the [O III] emission, indicating that the 8.7 \micron~emission is not only tracing NLR gas and star formation, but also the oval
structure of $\sim$350 pc radius detected by \citet{Riffel11} in the NIR. In an attempt to quantify the contribution of the NLR
to the extended MIR emission, we performed a {\sc galfit} modelling of the 8.7 \micron~GTC/CC image, which is described 
in Appendix \ref{appendixB}. Unfortunately, because of the almost identical orientations of the NLR and the oval structure 
detected by \citet{Riffel11} in the NIR (135\degr~and 128\degr~respectively), we cannot separate the two components, as both 
are included in the Sersic profile used for the fit. We can, however, estimate an upper limit to the luminosity of the 
NLR by using the integrated flux of the {\sc galfit} model: Log L$_{NLR}\leq$ 43.20$\pm$0.07 erg~s$^{-1}$.

Finally, regarding the distribution of the molecular gas in Mrk\,1066, as traced by the NIR H$_2$ lines, 
\citet{Riffel10} found it
to be more conspicuous at the positions of the nucleus and knot A, and likely produced by X-ray heating of the circumnuclear gas, 
with some contribution from thermal emission. The influence of shocks in the H$_2$ emission can be quantified using 
the H$_2$/PAH ratio \citep{Ogle10}. Using the emission line fluxes reported by \citet{Riffel10} for knots N, A and B, and those 
measured from the PSF-subtracted 8.7 \micron~GTC/CC image using the same aperture (0.25\arcsec~diameter), we can study possible 
variations of the H$_2$/PAH ratio in the different knots. We find a larger ratio in knot A: 
H$_2$/PAH=(1.0$\pm$0.2)$\times10^{-3}$, than in knots N and B: H$_2$/PAH=(2.0$\pm$0.3)$\times10^{-4}$ and 
(5$\pm$1)$\times10^{-4}$ respectively. This is likely related to the radio jet interaction with the ISM, more 
important towards the NW.

\subsection{AGN dilution of the 11.3 \micron~PAH feature}
\label{discussion2}

In Section \ref{extended_spectroscopy} we presented the spectra of knots N, A, B and D before and after subtracting 
the AGN contribution that we estimated using the method there described.

Before subtracting the AGN component, the EW of the 11.3 \micron~PAH feature is lower in the 
nucleus than in the knots (see Table \ref{tab3} and Figure \ref{fig11}). On the other hand, once we subract the AGN contribution, 
the EWs increase (see last column in Table \ref{tab3}), specially in the nucleus. This shows that 
the relatively low EWs measured before AGN subtraction 
are due to the increasing AGN continuum as we approach the position of the active nucleus, rather than to PAH destruction.
We refer the reader to \citet{Alonso14} for further discussion on AGN dilution of the 11.3 \micron~PAH feature. 

The largest increase in EW corresponds to the nucleus (see Figure \ref{fig11}), where AGN dilution is higher. The EWs measured
for knots A and B, which are at distances $\leq$100 pc from the nucleus, also experience an increase. Finally, in the case
of knot D, at 200 pc NW to the nucleus, AGN dilution is almost negligible, and the EW remains the same within the errors. 
The EWs measured, after AGN subtraction, for knots A, B and D are lower than those reported by \citet{Alonso14} regions 
at 200--300 pc SE and NW of the nucleus (1.4--1.5$\pm$0.3 \micron). This is likely due to dilution produced by the
NLR emission. Note that here we are just subtracting the unresolved AGN emission, which dominates in the 
nucleus, but represents $\leq$36\% in the knots. We know from the results presented in \citet{Riffel10} and this work that 
there is extended IR emission from dust in the NLR of Mrk\,1066, and this emission might be
diluting the PAH features in the knots. 

Taking advantage of the spatial information afforded by the GTC/CC spectroscopy, we have shown 
that at least in the case of Mrk\,1066, the AGN continuum dilutes, rather than destroys, the PAH features 
on nuclear scales ($\sim$60 pc) 
and up to $\sim$100 pc from the active nucleus. This result is in agreement with
the findings of \citet{Alonso14} and \citet{Esquej14}. 


\section{Conclusions}

We present new MIR imaging and spectroscopic observations of the Sy2 galaxy Mrk\,1066 at subarcsecond 
angular resolution obtained with CanariCam on the 10.4 m GTC. The data probe the central $\sim$400 pc 
of the galaxy with an angular resolution of 54 pc, revealing a series of star-forming knots after 
subtracting the dominant AGN contribution from the MIR emission. Our major conclusions can be summarized as follows:

\begin{itemize}

\item By subtracting the dominant AGN contribution to the GTC/CC nuclear spectrum of Mrk\,1066, 
we find that the EW of the 11.3 \micron~PAH feature is larger in the nucleus than in the knots, with all of them
being typical of starburst galaxies. This confirms that, at least in the case of this galaxy, the AGN does not 
destroy, but dilutes, the molecules
responsible for the 11.3 \micron~PAH emission in the inner $\sim$60 pc of the galaxy and up to $\sim$100 pc from the nucleus. 

\item We measured the flux of the 11.3 \micron~PAH band in the knots before and after subtracting the 
AGN contribution to the nuclear spectra, and we find SFRs = 0.2--0.3 M$_\odot~yr^{-1}$. These values coincide with the 
largest values measured by \citet{Esquej14} for local Seyfert galaxies in regions of $\sim$65 pc in size. 


\item We fitted the nuclear NIR and MIR SED of Mrk\,1066 with clumpy torus models, and derived a torus gas mass
of 2$\times10^5~M_{\odot}$, contained in a clumpy torus of $\sim$2 pc radius. Besides, we derived a column 
density from the fit that is compatible with 
Mrk\,1066 being a Compton-thick candidate, in agreement with X-ray observations.

\item By comparing the nuclear GTC/CC MIR spectrum with the \emph{Spitzer}/IRS spectrum of Mrk\,1066, and performing spectral 
decomposition into AGN and starburst components, we find that the AGN component that dominates the continuum emission at $\lambda<$15 
\micron~on scales of $\sim$60 pc (90--100\%), decreases to 35--50\% when the emission of the central $\sim$830 pc is considered.

\item The AGN contribution to the \emph{Spitzer}/IRS 3.7\arcsec~spectrum dominates the MIR emission at 15--25 \micron~(75\%), which is 
among the highest percentages measured for local LIRGs using the same methodology.  


\item We find a good match between the MIR morphology and the extended Pa$\beta$, Br$\gamma$ and [O III]$\lambda$5007 emission. 
This coincidence implies that the 8.7 \micron~extended emission in Mrk\,1066 is probing star formation, dust in the NLR, and 
also the oval structure previously detected in the NIR. 

\item The \emph{Chandra} soft X-ray morphology does not match either the IR or the [O III]$\lambda$5007 extended emission, 
implying that it is not tracing dust in the NLR. 
Instead, the multiwavelength data analyzed here favour a thermal origin for the soft X-ray emission.

\end{itemize}

\appendix

\section{Bayesian inference.}
\label{appendixA}

\begin{figure}
\includegraphics[width=9cm]{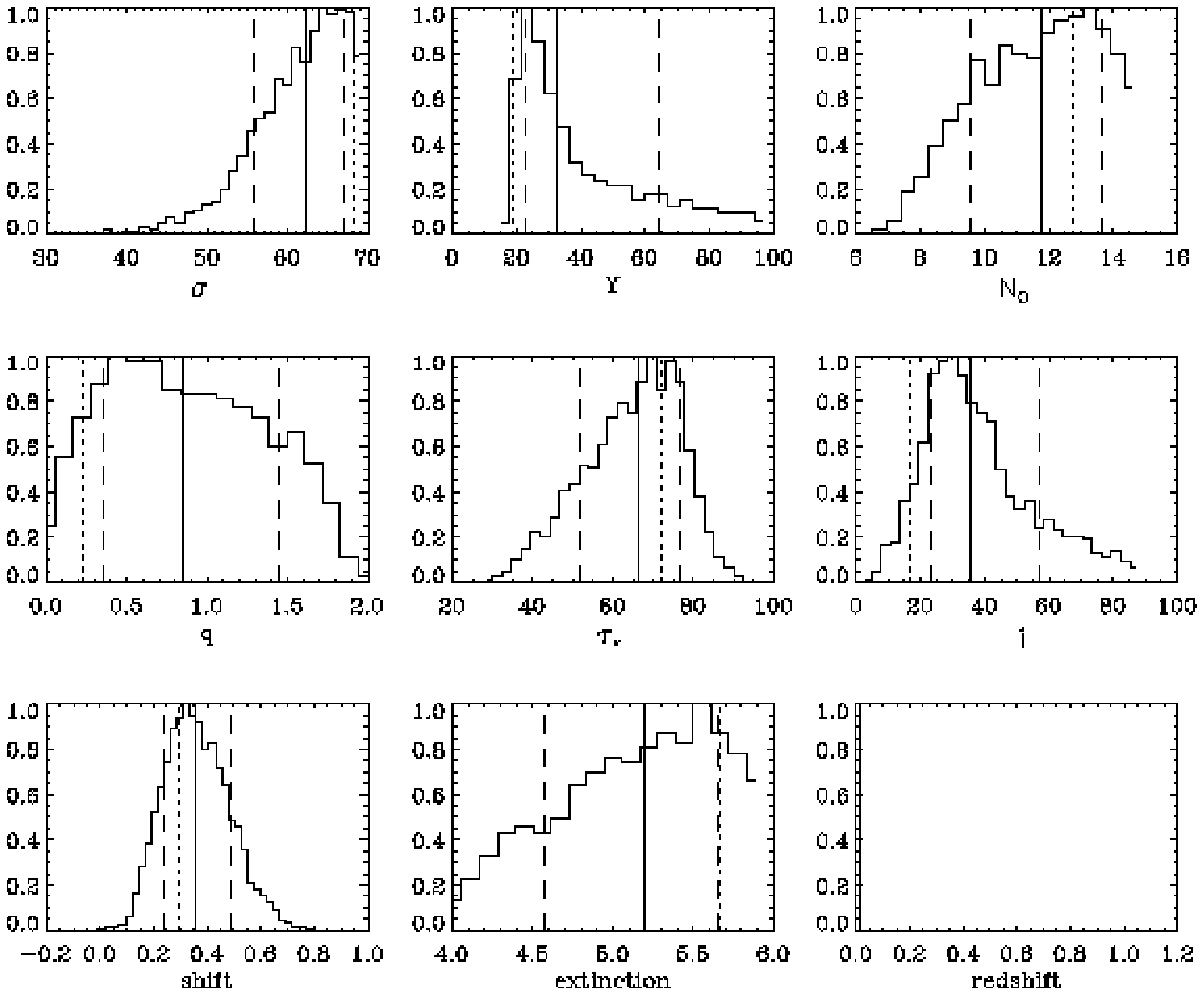}
\caption{Normalized marginal posteriors resulting from the fit of Mrk\,1066 nuclear IR SED. 
Dotted and solid vertical lines represent the MAP and the median of each posterior, respectively, 
and dashed vertical lines indicate the 68 per cent confidence level for each parameter around the median 
(see Table \ref{tab2}). 
\label{A1}}
\end{figure}

Here we report the posterior distributions resulting from the fit of the IR SED of Mrk\,1066. 
The top and middle rows of Figure \ref{A1} correspond to the posteriors of the six parameters that describe the models 
(defined in Table \ref{tab2}). The bottom row includes the posteriors of the foreground extinction, A$_V$,
the vertical shift required to match the fluxes of a chosen model to an observed SED, and the galaxy redshift.

\section{Galfit Modelling.}
\label{appendixB}

In order to better characterize the MIR morphology of Mrk\,1066, we performed {\sc galfit} 
(\citealt{Peng02,Peng10}; version 3.0.5) modelling of the 8.7 \micron~GTC/CC image.
{\sc galfit} is a well-documented two-dimensional fitting algorithm which allows 
the user to simultaneously fit a galaxy image with an arbitrary number 
of different model components, and thus to extract structural parameters of the galaxy. 
The model galaxy is convolved with a PSF and, using the 
downhill-gradient Levenberg-Marquardt algorithm, is matched to the observational 
data via the minimization of the $\chi^2$ statistics.

We used the image of the standard star as PSF component, and the host galaxy was modelled using 
a Sersic profile to account for the oval structure of $\sim$350 pc detected by \citet{Riffel11} in the NIR. 
All the model parameters were allowed to vary freely except the PSF flux, which we fixed to be the same 
we obtained from PSF subtraction (63$\pm$9 mJy), and the position angle of the Sersic component (128\degr; \citealt{Riffel11}). 
We fixed the flux of the PSF component
because otherwise the residual image appeared clearly oversubtracted at the nuclear position. After 
doing so, the best fit resulted in a Sersic profile with index n=0.83 (i.e., consistent with a disk), effective radius 
R$_{eff}$=265 pc and ellipticity b/a=0.59. The final reduced-$\chi^2$ value is 1.156.

\begin{figure*}
\centering
\includegraphics[width=17cm]{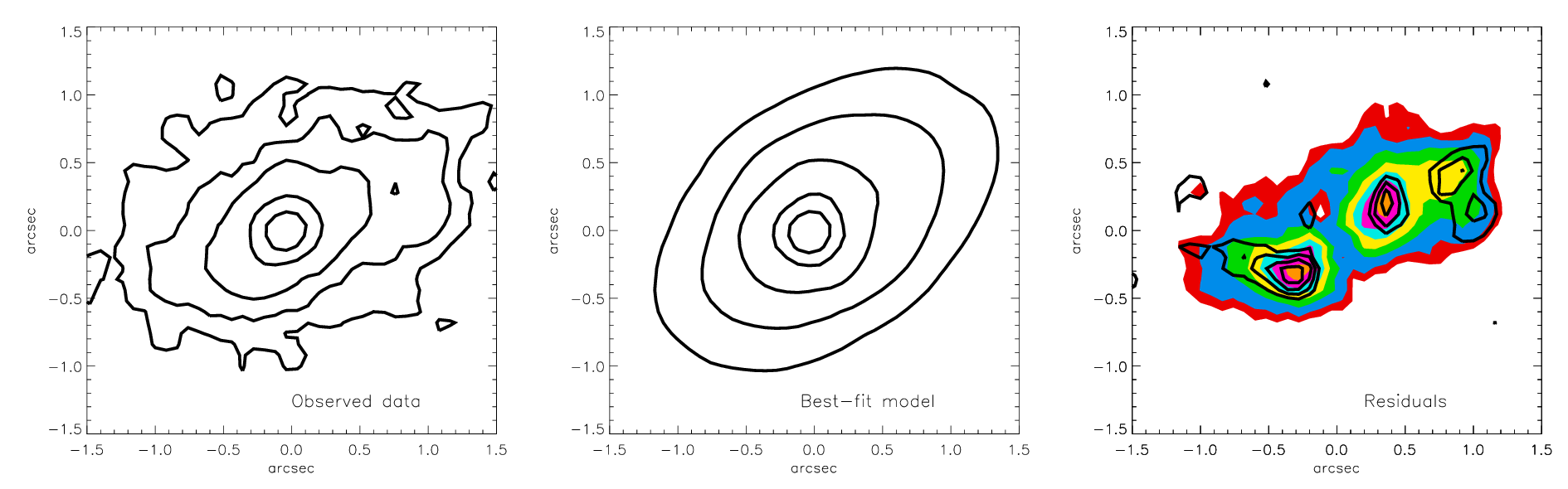}
\caption{From left to right: contours of the 8.7 \micron~GTC/CC image, the best-fitting model and the model-subtracted 
residual image. The residuals are overlaid on a colour-scale of the PSF-subtracted MIR image, for comparison.
\label{B1}}
\end{figure*}

In Figure \ref{B1} we display the 8.7 \micron~GTC/CC contour plots of the central 670$\times$670 kpc$^2$ region of Mrk\,1066,
the best-fitting model and the model-subtracted residual image. The 
residuals are overlaid on a colour-scale of the PSF-subtracted MIR image, for comparison. Only the star-forming 
knots appear in the residual, as the rest of the MIR extended emission is included in the disk model. 
If we measure the flux of the residual in an aperture of 2\arcsec~radius, we obtain 2.5$\pm$0.4 mJy. This is indicating that
the Sersic component is likely including not only the disk emission but also the NLR emission, with the latter traced by 
the [O III] image shown in Figure \ref{fig10}. This is likely due
to the almost identical orientations of the NLR and the disk (135\degr~and 128\degr, respectively). Although we cannot
separate the two components in the MIR, we can estimate an upper limit to the NLR emission by using the integrated 
flux of the {\sc galfit} model\footnote{The integrated flux of the model does not include either the nuclear component 
or the star-forming knots.}, which is 172$\pm$26 mJy. 

\section*{Acknowledgments}
This research was supported by a Marie Curie Intra European Fellowship
within the 7th European Community Framework Programme (PIEF-GA-2012-327934).
CRA and IGB ackowledge financial support from the Instituto de Astrof\' isica de Canarias and the 
Spanish Ministry of Science and Innovation (MICINN) through project 
PN AYA2010-21887-C04.04 (Estallidos). IGB is grateful to the PhD contract funded by Fundaci\'on La Caixa.
AAH acknowledges support from the Spanish Plan Nacional de Astronom\' ia y Astrof\' isica under grant
AYA2012-31447 and from the Augusto G. Linares Program through the Universidad de Cantabria.
RAR thanks the support of Brazilian institutions CNPq and FAPERGS.
PE acknowledges support from the Spanish Programa Nacional de Astronom\' ia y Astrof\' isica under grant AYA2012-31277.
CP acknowledges support from UTSA to help enable this research.
OGM and JRE ackowledge financial support from the 
Spanish Ministry of Science and Innovation (MICINN) through project AYA2012-39168-C03-01.

Based on observations made with the Gran Telescopio CANARIAS (GTC), installed in the Spanish Observatorio
del Roque de los Muchachos of the Instituto de Astrof\' isica de Canarias, in the island of La Palma. 

This research has made use of the NASA/IPAC Extragalactic Database (NED) which is operated by the Jet Propulsion Laboratory,
California Institute of Technology, under contract with the National Aeronautics and Space Administration. 

The scientific results reported in this article are based in part on data obtained from the \emph{Chandra} Data Archive. 

Based on observations made with the NASA/ESA Hubble Space Telescope, obtained from the data archive at the 
Space Telescope Science Institute. STScI is operated by the Association of Universities for Research in Astronomy, 
Inc. under NASA contract NAS 5-26555.

Based on observations obtained at the Gemini Observatory, which is operated by the Association of Universities 
for Research in Astronomy, Inc., under a cooperative agreement
with the NSF on behalf of the Gemini partnership: the National Science Foundation (United
States), the Science and Technology Facilities Council (United Kingdom), the
National Research Council (Canada), CONICYT (Chile), the Australian Research Council
(Australia), Minist\'erio da Ci\^encia e Tecnologia (Brazil)
and Ministerio de Ciencia, Tecnologia e Innovaci\'on Productiva  (Argentina). 

The authors acknowledge Santiago Garc\' ia Burillo, Andr\'es Asensio Ramos and Mar Mezcua for useful discussions.
Finally, we are extremely grateful to the GTC staff for their constant and enthusiastic support, 
and to the anonymous referee for useful comments.

\label{lastpage}

\end{document}